\documentclass[nofootinbib,aps,prd,preprint, superscriptaddress,
showpacs,amssymb]{revtex4}

\usepackage{amsmath,amsfonts,amssymb}

\usepackage{graphics}
\usepackage{color}
\usepackage{dcolumn}
\usepackage{dcolumn,epsfig}

\allowdisplaybreaks

\newcommand{\be}{\begin{equation}}
\newcommand{\ee}{\end{equation}}
\newcommand{\bea}{\begin{eqnarray}}
\newcommand{\eea}{\end{eqnarray}}

\newcommand{\ben}{\begin{eqnarray*}}
\newcommand{\een}{\end{eqnarray*}}
\newcommand{\bs}{\begin{subequations}}
\newcommand{\es}{\end{subequations}}

\begin{document}

\title{Gravitational Recoil during Binary Black Hole Coalescence using 
the Effective One Body Approach }

\date{\today}

\author{Thibault Damour}
\affiliation{
Institut des Hautes Etudes Scientifiques,
91440 Bures-sur-Yvette, France}

\author{Achamveedu Gopakumar}
\affiliation{
Theoretisch-Physikalisches Institut,
Friedrich-Schiller-Universit\"at, Max-Wien-Platz 1, 07743 Jena, Germany
}

\begin{abstract}

During the coalescence of binary black holes, gravitational waves
carry linear momentum away from the source, which results in the recoil of
the center of mass. Using the Effective One Body approach, that includes
nonperturbative resummed estimates for the damping and conservative
parts of the compact binary dynamics, we compute the recoil
during the late inspiral and the subsequent plunge of non-spinning
black holes of comparable masses moving in quasi-circular orbits.
Further, using a prescription that smoothly connects the plunge phase
to a perturbed single black hole, we obtain an estimate
for the total recoil associated with the binary black hole
coalescence. We show that the crucial physical feature which determines
the magnitude of the terminal recoil is the presence of a ``burst''
of linear momentum flux emitted slightly before coalescence.
When using the most natural expression for the linear momentum
flux during the plunge,
together with a Taylor-expanded $(v/c)^4$ correction factor,
we find that the maximum value of the terminal recoil is
$\sim 74$ km/s and occurs for $\eta = \frac{ m_1\, m_2} { (m_1+ m_2)^2}  \simeq 0.2$,
{i.e.}, for a mass ratio $m_2/m_1 \simeq 0.38$.
Away  from this optimal mass ratio,
the recoil velocity decreases approximately
proportionally to the scaling function
$\tilde f (\eta) = \eta^2\, 
\left ( 1 - 4\, \eta \right )^{1/2}\, \left ( 1.0912 - 1.04\,\eta + 2.92\, \eta^2 \right ) $.
We comment, however, on the fact that the above
`best bet estimate' is subject to strong uncertainties
because the location and amplitude of the crucial peak of linear
momentum flux happens at a moment during the plunge where most
of the simplifying analytical assumptions underlying the
Effective One Body approach are no longer justified.
Changing the analytical way of estimating the linear momentum flux,
we find maximum recoils that range between 49 and 172 km/s.

\end{abstract}

\pacs{04.30Db, 04.25.Nx, 04.80.Nn, 95.55.Ym}

\maketitle

\section{Introduction }\label{secI}

 During the coalescence of compact binaries, along with energy and angular 
momentum, the system radiates linear momentum.
The loss of linear momentum via gravitational radiation results in the recoil
of the center of mass of the binary.
It is astrophysically important and desirable to obtain a dependable
estimate for the velocity of the center of mass of comparable-mass
black hole binaries undergoing coalescence \citep{RR89,MMFHH04}.
Notably, in models of massive black hole formation involving successive
mergers, recoils large enough to eject coalescing black holes from dwarf galaxies 
or globular clusters would effectively terminate the process. This motivation has 
recently led several authors to estimate the recoil velocity of coalescing black hole 
binaries by means of various methods \citep{FHH04,MC05,BQW05}.
Ref.~\cite{FHH04} employed black hole perturbation theory to describe the 
motion of a test mass moving in a black hole background, {i.e.},
the case where the symmetric mass ratio $\eta= \frac{ m_1\, m_2} { (m_1+ m_2)^2} \ll 1 $.
They combined a numerical estimate of the recoil velocity up to 
the Last Stable Orbit (LSO), with two crude estimates for the recoil acquired during the
subsequent plunge phase. Then they assumed that their test-mass estimates could be
proportionally scaled up to the comparable-mass cases ($4\, \eta \sim 1$)
with the function
\be
f(\eta) = \eta^2\, \sqrt{ 1- 4\, \eta}\,,
\label{feta}
\ee
which appears as an overall factor in the leading (``Newtonian")
analytical estimate of the recoil, as first computed in Ref.~\cite{F83}.
The final estimates of Ref.~\cite{FHH04} range (for non-spinning black holes) between a
lower value $\sim 54 \, \left( f(\eta)/f_{\rm max} \right ) \, {\rm km/s} $ and
an upper one $\sim 465\, \left( f(\eta)/f_{\rm max} \right ) \, {\rm km/s} $,
where $f_{\rm max} = f(\eta_{\rm max}) = f(0.2) = 0.0178885 $.
Another set of estimates was obtained from an approach that employs a mixture of 
numerical relativity and black hole perturbation theory for 
the merger of comparable mass non-spinning black holes 
\citep{MC05}. In particular, for a mass ratio $m_2/m_1 =0.5$,  corresponding to
a value $\eta = 0.2222$, close to the value $\eta_{\rm max} =0.2$, where 
$f(\eta)$, given by Eq.~(\ref{feta}), reaches its maximum value
$ [  f (0.2222)/f(0.2) = 0.01646/0.01789 \simeq 0.92 ]$,
Ref.~\cite{MC05} estimates a recoil velocity $\sim 250 \pm 150 \,{\rm km/s} $.
Finally, 
using an analytical estimate, which is further discussed below,
a maximum recoil (reached for $\eta \sim 0.2$ ) equal to
$ 250 \pm 50 \,{\rm km/s} $
was obtained in Ref.~\cite{BQW05}.

Summarizing, the recent estimates
are consistent with a maximum recoil velocity
$\sim 250\, {\rm km/s} $ for non-spinning black holes
\citep{FHH04,MC05,BQW05}.
In contrast, we shall
estimate here, by using the Effective One Body (EOB) approach to binary black hole
dynamics, detailed in Refs.~\citep{BD99,BD00,DJS00,TD01,BCD05}, that the maximum recoil velocity
for non-spinning coalescing black holes is probably 
significantly smaller, and of the order $\sim 50-74\,{\rm km/s} $. However, we shall conclude that this estimate is rather uncertain because it depends on
a specific way of describing the linear momentum flux during
a crucial phase of the plunge which is (mildly) relativistic,
and has not been yet analytically studied in detail.
When changing our preferred asssumptions for describing the
linear momentum flux, we find maximal recoil velocities that vary in the
range $49-172$ km/s.

 Let us recall that the coalescence of isolated 
black hole binaries may be viewed  as consisting  of the following
three phases. 
The first phase is that of gravitational-radiation-driven slow inspiral
in quasi-circular orbits. This leads, after a very long period of
adiabatic shrinkage of the orbital separation, to the binary
approaching its
LSO.  
At the LSO, the inspiral phase changes to some sort of plunge, dominated by  general relativistic strong-field effects.
This plunge phase results in the merger of the two black holes and
the dynamics of the final black hole, thus formed,  may be described in terms 
of black hole quasi-normal modes.
 
 During the early inspiral, the dynamics of compact binaries
can be described very accurately by the post--Newtonian (PN) approximation to general relativity.
The PN approximation to general relativity
allows one to express the equations of motion for a compact binary
as corrections to Newtonian equations of motion
in powers of $(v/c)^2 \sim GM/(c^2 R)$, where $v, M,$ and $R$ are
the characteristic orbital
velocity, the total mass and the typical orbital separation of the binary respectively.
For a compact binary, treated to consist of non-spinning point masses,
of arbitrary mass ratio $\eta = \frac{m_1\, m_2}{M^2} $,
where $m_1$ and $m_2$ are the masses of the components and $M= m_1 + m_2$, the leading
``Newtonian"  contribution \footnote{Or rather ``quasi-Newtonian", as this leading contribution
corresponds to 3.5PN  $ { \cal O}(v^7/c^7) $] radiation reaction terms
in the equations of motion for the binary.}
to linear momentum flux $ \mathcal{F}^i_\mathbf{P}$ and 
the associated instantaneous recoil ${\mathbf v}^i_{\rm com}$,
{i.e.}, the instantaneous
velocity of center of mass can be derived from 
the investigations that 
dealt either with wave-zone flux computations, or 
with near-zone radiation-reaction ones
\citep{BR61,Pe62,Pa62,KT80}.
This allowed Ref.~\cite{F83} to derive the following explicit leading order results
(for a circularized orbit of radius $R$)
\bs
\label{N_estimates}
\bea
| \mathcal{F}^i_\mathbf{P}| &=&  \frac{ 464\, c^4}{105\, G}\, 
f(\eta)\,
\left ( \frac{ G\,M }{c^2\, R} \right )^{11/2}\,,\\
| \frac{v^i_{\rm com} }{c}| &=&  \frac{464}{105}\, 
f(\eta)\,
\left ( \frac{ G\,M }{c^2\, R} \right )^{4}\,.
\eea
\es
where $f(\eta)$ denotes the combination introduced in Eq.~(\ref{feta}).
The function $f(\eta)$ vanishes both in the limit of extreme mass ratio (test-mass limit,
$\eta \rightarrow 0$), and in the case of equal mass binaries ($\eta = 1/4$).
It reaches the maximum value
$f_{\rm max} = 0.0178885$
for $\eta_{\rm max} =0.2$,
corresponding to a mass ratio $m_2/m_1 = \left ( 3 \pm \sqrt{5} \right )/2 \sim 0.3820$
or $2.618$. In the following, we shall generally
(though not systematically) employ units where $G=c=1$.

  The above leading-order results do not allow one to reliably estimate the final recoil of
a coalescing binary black hole. Indeed, on the one hand they neglect higher-order PN corrections
that might become fractionally large near the LSO and during the subsequent plunge,
and on the other hand, they do not take into account the crucial transition between 
quasi-circular inspiral and plunge. As exemplified by previous
works \cite{FD84,FHH04,MC05,BQW05},
one can think of several different ways of going beyond the above results,
given by Eqs.~(\ref{N_estimates}). As first attempted by
Detweiler and Fitchett \cite{FD84},
one can use perturbation theory
around black hole backgrounds to calculate the linear momentum
flux emitted by a test particle moving around a black hole,
assuming then a scaling proportional to $f(\eta)$ in order to go from the test-mass case
($\eta \ll 1)$ to the comparable-mass one. 
Ref.~\cite{FD84} considered only circular motions (both above and below LSO), while 
Ref.~\cite{FHH04} combined information about circular orbits above the LSO with crude estimates
of the linear momentum flux during the plunge following the LSO crossing.
However, we note that this approach is yet to be applied for the relevant case of 
(non-geodesic) plunging motions.
The Lazarus program, which employs a mixture of perturbation theory and numerical relativity,
to study the late inspiral and the merger of black hole binaries
is not limited to the case $\eta \ll 1$. However, these simulations are limited to rather
short evolution time spans. For instance, Ref.~\cite{MC05} mentions that the simulations are
accurate only for ``less than 15 $M$ in time", which is significantly smaller than the 3PN-accurate
estimate of the orbital period of a comparable-mass black hole binary
at the LSO: $T^{\rm 3PN}_{\rm LSO} \simeq 71.2\, M $ \citep{DJS00}.
This forces state-of-the-art numerical simulations to employ initial data sets where
the two holes are already quite near to each other (and, in fact, probably closer than the LSO).
It is indeed a challenge to construct initial data sets for tight black hole binaries,
and which do correspond to the physically correct ``no-incoming radiation" criteria.
[See Refs.~\cite{BGGN04,SG04} for prescriptions to construct `realistic initial data sets'.]
Presence of spurious radiation in the initial data sets is likely to dominate 
and thereby invalidate the estimate of the linear momentum flux.
However, a  comparison between numerically constructed initial data sets and analytically
determined characteristics of tight binary systems, performed in Ref.~\citep{DGG}, have
indicated the ability of the Effective One Body (EOB) approach 
\citep{BD99,DJS00,TD01} to describe rather accurately the initial data
obtained by the helical Killing vector method \citep{GGB}, which is carefully
aimed at minimizing the amount of spurious radiation present in the initial state.

   In view of the above situation, we consider that our present ``best bet" for obtaining a 
dependable estimate for the gravitational recoil during the late inspiral and plunge phases 
of a black hole binary consists in employing the EOB approach.
This approach has already been used to compute complete  gravitational waveforms 
emitted during the inspiral and merger of (non-spinning and spinning) black holes
\citep{BD00,BCD05}.
As first proposed in Ref.~\cite{BD00}, this is achieved by considering the waveform emitted by the 
binary beyond the LSO, through the subsequent ``plunge", down to, approximately, the 
``light ring" ($R \simeq 3\,M$), and by matching it there to a ``ring down" signal 
constructed using the quasi-normal modes of the resultant final black hole.
As this (extended) EOB approach will be central to the present paper, let us recall
the main arguments of Ref.~\cite{BD00} for proposing the apparently bold strategy
of analytically describing the plunge beyond the LSO, down to $R\simeq 3\,M$.
A first argument is that the EOB approach is a resummation technique which was 
carefully devised to work not with the usually considered, badly convergent, PN-expanded 
equations of motion or flux quantities, but instead with a (EOB) re-summed Hamiltonian
and a (Pad\'e) re-summed damping force showing no sign of bad behavior during most of 
the ``plunge". In particular, it was found in Ref.~\cite{BD00} that the word ``plunge"
to qualify the dynamics beyond the LSO is a misnomer, and that this phase is better 
thought of as being still a quasi-circular inspiral motion, even down to the 
light ring $ R \simeq 3\,M$. Indeed, it was found that the quasi-circularity
condition ($ \dot{R} \ll R \dot{\varphi}$)
remains satisfied with good accuracy beyond the LSO, down to
$ R \simeq 3\,M$.
This is illustrated in Fig.~\ref{RKE_AKE_fig} which shows (for $\eta = 0.2$)
the evolution of the ``azimuthal'' ($ g^{\varphi \varphi} p_{\varphi}^2$)
and  ``radial'' ( $ g^{RR} p_R^2$)  kinetic energies
during the plunge down to the light-ring, $R \simeq 3\,M$.
The crucial point is that the ratio 
$ {\cal R} \equiv g^{RR} p_R^2 / g^{\varphi \varphi} p_{\varphi}^2$
stays significantly smaller than one during the entire plunge.  Its
value at $R \simeq  3\,M $ is ${\cal R}_{\rm light-ring} \simeq 0.281 $.

As for the idea of matching the gravitational wave emission to a quasi-normal-mode (QNM)
``ring-down" signal around $R\simeq 3\,M $, let us recall that it was realized long ago
that the basic physical reason underlying the presence of a
QNM-type merger signal, that ends the plunge signal, was that the $ l \ge 2$
gravitational waves emitted by the
collapsing system are strongly filtered by the potential barrier, centered around
$R \simeq 3\,M $, describing the radial propagation of the 
gravitational waves
\citep{DRPP71,DRT72,P71}
\footnote{For the test-particle case, this follows
from the explicit form of the Regge-Wheeler-Zerilli effective potentials.
In the comparable-mass case, we must contemplate the $l \ge 2$ binary gravitational
wave signal as propagating in the spacetime generated by the binary system, 
and approximate the latter (when the two holes are closer than $3\,M$, and when 
considering the waves propagating in the domain $R \ge 3\, M$) by the external 
geometry of a single hole (of mass-energy $\simeq M$).}.

  Recently, Blanchet, Qusailah and Will~\cite{BQW05} have employed an ``approximation" to the EOB method
 in the sense that they `assume that the plunge can be viewed
as that of a ``test particle" of mass $\mu = \frac{m_1\, m_2}{M} $ moving in the
fixed Schwarzschild geometry of a body of mass $M$'.
They also assumed that the effect of radiation reaction damping on the plunge 
orbit may be ignored. They then matched, in various ways, a circular orbit at the 
Schwarzschild LSO, {i.e.}, $6\,M$, to a suitable plunge orbit. By contrast,
we shall use here the {\em full EOB} approach \cite{BD00}, which does not 
need to ``assume" that the plunge can be viewed as that of suitable test particle, but instead
proves it [see Refs.~\citep{BD99,BD00,DJS00}], and which does not need to match a circular orbit
to a plunge orbit at the LSO, because it automatically embodies 
a {\it smooth transition} between the ``inspiral" and the ``plunge".
Let us also note that the ``effective test body" used in the EOB method does not evolve
in a fixed Schwarzschild geometry of mass $M$, but instead in a deformed Schwarzschild
background, whose geometry was algorithmically derived to 2PN accuracy in Ref.~\cite{BD99},
and to 3PN order in Ref.~\cite{DJS00} [see also Ref.~\cite{TD01} for
the incorporation of spin effects].
In addition, while Ref.~\cite{BQW05} formally let their ``test particle"  fall down to the
horizon at $ R = 2\,M$, an important ingredient of our approach will be to match
the plunge signal to a QNM-based ring-down one at $R \simeq 3\,M$.

  On the other hand, an important result of Ref.~\cite{BQW05}  concerns the higher-order PN
corrections to the ``Newtonian" linear momentum flux, given by Eq.~(\ref{N_estimates}).
Using the multi-polar post-Minkowskian approach \citep{BD86,BD89,DI91,BD92,LB95,LB98},
and its higher-order implementations \citep{BIJ2002,BDEI04,ABIQ04},
Ref.~\cite{BQW05} has gone beyond the previous 1PN-accurate studies of recoil
effects, available in Ref.~\citep{W92}, by including both the 1.5PN order ``tail" contribution
and the next 2PN order corrections. 
Ref.~\cite{BQW05}
finds that the linear  momentum flux at infinity, for binary systems in {\em circular orbits},
is given by a PN expansion of the form
\be
\mathcal{F}^{i{({\rm BQW})} }_\mathbf{P}
= -\frac{ 464\,}{105\,} \, f(\eta)\, v_{\omega}^{11}\, F( v_{\omega}; \eta) \,\lambda^i\,,
\label{BQWflux}
\ee
where $f(\eta)$ is given by Eq.~(\ref{feta}) above 
\footnote{We conventionally assume henceforth that $m_1 \ge m_2$ so that 
$\frac{ m_1 -m_2}{M} = + \sqrt{1- 4\, \eta} $.},
and 
\be
v_{\omega} \equiv \left ( M\, \Omega \right )^{1/3}\,,
\label{vomega}
\ee
with $\Omega \equiv d \varphi/dT$  denoting the orbital angular velocity.
The factor  $F( v_{\omega}; \eta)$ yields the 2PN-accurate ``Taylor-expanded"
PN-corrections to the linear momentum flux 
(when the latter is expressed in terms of the above defined $v_{\omega}$)
and it reads
\be
\label{FTaylor}
F( v_{\omega}; \eta) = 1 + F_2(\eta)\, v_{\omega}^2 + F_3 (\eta)\, v_{\omega}^3
+ F_4 (\eta)\, v_{\omega}^4\,,
\ee
with
\bs
\bea
F_2(\eta) &=& -\frac{452}{87} -\frac{1139}{522}\eta \,,
\\
F_3 (\eta) &=& \frac{309}{58}\,\pi\,,
\\
F_4(\eta) &=& -\frac{71345}{22968}+\frac{36761}{2088}\eta
+\frac{147101}{68904}\eta^2 \,.
\eea
\es
Finally, $\lambda^i$ in Eq.~(\ref{BQWflux}) is a tangential unit vector directed in the same sense
as the relative orbital velocity $v^i \equiv v_1^i - v_2^i$,
$v_1^i$ and $v_2^i$  being the velocities of the masses $m_1$ and $m_2$
respectively.
Note that the test-mass limit ($\eta \rightarrow 0 $) of the function $F(v_{\omega})$
has been first numerically evaluated in Ref.~\cite{FD84}.
As we shall see below, one of the important differences between our
treatment and the one of Ref.~\cite{BQW05} will concern the continuation
of the linear momentum flux Eq.~(\ref{BQWflux}) (derived for circular
orbits above the LSO) to the (non circular)
plunging orbit below the LSO.

  In the next section, we present our prescription to compute 
the  linear momentum flux and the related velocity of center of mass. 
Section~\ref{S3} contains a  summary of the `modified' EOB
approach that is used to describe the late stages of binary inspiral and plunge,  
followed by a detailed account of the numerical procedure that will result in the determination 
of the associated gravitational radiation driven recoil.
We also present in that section analytical insights into our numerical estimates.
In Sec.~\ref{S4}, we describe how we smoothly match the merger and the resultant ring down
phases and the computation of the recoil of the final black hole. 
We present our results, conclusions and future directions in Sec.~\ref{S5}.

\section{Quasi-Newtonian formulas for linear momentum flux and related recoil}
\label{S2}

 In the EOB formalism, one finds that the relative orbital dynamics of a binary black hole
system is most conveniently described in a ``Schwarzschild-like" coordinate system, to 
which is associated an ``effective metric" of the form
\be
\label{geffective}
ds^2_{\rm eff} = - A(R)\, dT^2_{\rm eff} + \frac{D(R)}{A(R)}\, dR^2 
+ R^2\, \left ( d \theta^2 + \sin^2 \theta\, d \varphi^2 \right )\,.
\ee
We shall work here to 2PN accuracy, in which case the ``effective metric coefficients"
$A(R)$ and $D(R)$ are given by
\bs
\bea
\label{coefA}
A(R) &=& 1 - \frac{2\,M}{R} + 2\,\eta \, \left( \frac{M}{R} \right )^3\,,
\\
\label{coefD}
D(R)&=& 1 - 6\, \eta \, \left( \frac{M}{R} \right )^2\,.
\eea
\label{eff_A_D}
\es
 It was shown in Ref.~\citep{BD99}
 that the complicated and badly convergent second post-Newtonian
expanded dynamics of a binary system could be mapped onto the much simpler 
(and better convergent) dynamics of an auxiliary test particle falling along
a geodesic of the effective metric, Eq.~(\ref{geffective}). Note that,
even in the equal mass limit ($\eta = \frac{1}{4}$), the effective metric
coefficients, given by 
Eqs.~(\ref{eff_A_D}), differ only slightly from those of a 
Schwarzschild metric ({i.e} $ A_{\rm S}(R) = 1- \frac{2\,M}{R}; D_{\rm S} (R) =1 $). As emphasized in Ref.~\citep{BD99} this property
makes it useful to describe the EOB dynamics in the Schwarzschild-type
coordinates of Eq.~(\ref{geffective}), rather than, say, in
Arnowitt-Deser-Misner or harmonic coordinates which would lead either
to a (badly convergent) infinite series of PN corrections, or to
more complicated ``resummed'' expressions.

  One of the important features of the present study will be
to express the flux of linear momentum
radiated  away from a compact binary directly in terms of the
quasi-Schwarzschild coordinates $R$ and $\varphi$
used in the effective one body metric, Eq.~(\ref{geffective}),
and of their time-derivatives, notably the angular velocity
$\Omega = \frac{d \varphi}{dT}$. Our work will often rely
on the use of quantities having simple (and ``quasi-Newtonian'')
expressions in terms of quasi-Schwarzschild coordinates $R$ and $\varphi$.
Let us first motivate this use of quasi-Newtonian quantities
expressed in quasi-Schwarzschild coordinates.

In the test-mass limit, it is a striking feature of
Schwarzschild coordinates that they often allow one to
convert Newtonian results into exact, or near-exact,
Einsteinian results. A famous example of that is the location
( \`a la Mitchell-Laplace) of the Schwarzschild horizon
which is correctly given by using
the purely Newtonian energy conservation: $c(r)^2/2 - G M/R = c(\infty)^2$.
In addition, the angular frequency along
circular geodesics in a Schwarzschild background
is described, in Schwarzschild coordinates, by the usual Kepler law:
$ GM = \Omega^2 R^3$, so that the linear velocity $ v \equiv \Omega R$
is given by the usual Newtonian formula $v^2 = G M/R $. Here we shall
use the remarkable fact that this closeness
extends to gravitational radiation properties. In particular, the
total energy flux emitted by circular geodesics into gravitational waves
is numerically very well approximated by the simple quasi-Newtonian
formula obtained by writing the leading-order quadrupole formula
[$ dE/dt \propto (d^3 I_{i j}/ dt^3)^2 \propto {\Omega}^6 (I_{ij})^2$]
in Schwarzschild coordinates. Indeed, this yields
$ dE/dt = \frac{32}{5} \eta^2 {\Omega}^6 R^2$ which is quite close to
the complete general relativistic answer \cite{CPFS93}:
even at the LSO,
$R_{\rm LSO} = 6 M$,
the quasi-Newtonian result $\frac{32}{5} \eta^2 {\Omega}^6 R^2$
is only 12\% smaller than the full Einsteinian  one, and the agreement
is better for orbits above the LSO.
A look at Fig.~2 in Ref.~\citep{FD84}
shows that a similar type of agreement holds also for the flux of
linear momentum down to the LSO. Note that is crucial in this
comparison to ``interpret'' quasi-Newtonian results in terms of
Schwarzschild coordinates. For instance, if one were to insert
harmonic coordinates $R_h$ in the
quadrupolar result $ dE/dt =\frac{32}{5} \eta^2 {\Omega}^6 R_h^2$
(and uses a corresponding harmonic-coordinate Kepler law
$ M = \Omega^2 R_h^3$) one would obtain an estimate for the
energy flux which would be larger than the correct one, at the LSO
($R_h =5$),
 by a factor $\simeq 2.19$.

 Regarding the plunging orbits below the LSO, another important feature
 of our treatment is that we do not
wish to insert in the leading-order,
``quasi-Newtonian'', energy and linear momentum fluxes the usually
assumed ``Kepler-type'' law relating the angular velocity
$\Omega$ to the radius $R$. Indeed, Kepler's law (which reads
$G M = {\Omega}^2 R^3$ in the test-mass limit $\eta \to 0$ and in
Schwarzschild coordinates)
is only valid, below the LSO, along the physically irrelevant
sequence of unstable circular orbits corresponding to a
{\em maximum} of the effective radial potential\footnote{Indeed, this  maximum of the
effective potential corresponds to an  unphysical angular momentum
$p_{\varphi} > p^{\rm LSO}_{\varphi}$. In contrast, the physically
relevant  plunge motion corresponds to
$p_{\varphi} < p^{\rm LSO}_{\varphi} $ and thereby to a particle gliding down a
flattish effective potential having no maximum (nor minimum) anymore,
{i.e}, a potential near but below the lowest radial potential plotted in Fig.~1 of Ref.~\cite{BD99}.}.
The ``violation'' of Kepler's law during
the plunge will be illustrated in Fig. 2 below.

With this motivation, let us derive from scratch the quasi-Newtonian
result for the linear momentum flux. We start from the following leading order formula,
available in Ref.~\cite{KT80},
\be
\mathcal{F}^x_\mathbf{P} + i\, \mathcal{F}^y_\mathbf{P} 
= \frac{ 1}{ 336\, \pi } \,
\biggl \{ 
\sqrt{ 14}\,  \stackrel{\!\!\!\!(3)}{I^{2-2} }\, \stackrel{\!\!\!\!(4)}{I^{31} }\,
+ \sqrt{ 210}\,  \stackrel{\!\!\!\!(3)}{I^{22} }\, \stackrel{\!\!\!\!(4)}{I^{3-3} }\,
- 14\,i\,  \stackrel{\!\!\!\!(3)}{I^{2-2} }\, \stackrel{\!\!\!\!(3)}{S^{21} }\,
\biggr \}\,,
\label{P_flux_f}
\ee
where $\mathcal{F}^x_\mathbf{P}$ and $\mathcal{F}^y_\mathbf{P}$ are the $x$ and $y$ components
of the linear momentum flux. Here $I^{lm}$ and $S^{lm}$ denote
 the ``mass'' and  ``spin'' (or ``current'')
 radiative multipole moments of the binary, while
$ \stackrel{\!\!\!\!(3)}{I^{lm} }$
 denotes the third time derivative of $I^{lm}$.
Under complex conjugation $I^{lm}$ and $S^{lm}$ transform as 
\be
I^{lm*} = (-1)^{m}\, I^{l-m}\,, \,\,\,\,\,\, S^{lm*} = (-1)^m\, S^{l-m}\,.
\label{cc_rules}
\ee

  We display below the relevant $I^{lm}$ and $S^{lm}$ required to compute 
the leading order contribution to  $\mathcal{F}^x_\mathbf{P} + \mathcal{F}^y_\mathbf{P}$ 
for compact binaries in circular orbits,
taken from Ref.~\cite{JS92},
\bs
\bea
\frac{ I^{22}}{M}  &=& \frac{2}{5}\, \sqrt{10\,\pi}\,\, \eta\, R^2\, e^{-2\,i\,\varphi}
\,,\,\,\, \,\,\,
\frac{ S^{21}}{M}  = - \frac{8}{15}\, \sqrt{10\,\pi}\,\, \eta\,
\sqrt{1-4\,\eta}\,\, R^3\, \Omega\, e^{-i\,\varphi}\,,
\\
\frac{ I^{31}}{M}  &=& - \frac{2}{105}\, \sqrt{35\,\pi}\,\, \eta\,
\sqrt{1-4\,\eta}\,\, R^3\, e^{-i\,\varphi}\,,
\,\,\,\,\,
\frac{ I^{33}}{M}  = \frac{2}{63}\, \sqrt{21\,\pi}\,\, \eta\, 
\sqrt{1-4\,\eta}\,\, R^3\, e^{-3\,i\,\varphi}\,,
\eea
\label{is_s_ip}
\es
where $\Omega = \frac{d \varphi}{dT}$.
The exact expression for, say, $  \stackrel{\!\!\!\!(3)}{I^{22} }
\propto \frac{ d^3  \left ( R^2\, 
e^{-2\,i\, \varphi} \right )}{ dT^3}  $ contains several terms 
proportional to
$ \frac{d^3 \varphi}{dT^3},
\frac{dR}{dT}\,\frac{d^2 \varphi}{dT^2},..\frac{d^3 R}{dT^3} $.
In the following, we consider an inspiralling and plunging relative orbit. For such an orbit,
the derivatives $\frac{d^n R}{dT^n}$ do not vanish. However,
as already mentioned above, it was pointed out in
Ref.~\cite{BD00} that even during the ``plunge" following the LSO crossing, the radial motion,
characterized by $\frac{dR}{dT}$, remained small compared to the azimuthal one 
$R\ \frac{d \varphi}{dT} $. We shall take advantage of this fact to
simplify the expression of the time-differentiated multipole
moments entering Eq.~(\ref{P_flux_f}) by keeping only the terms
proportional to  the time-derivatives of the azimuthal angle $\varphi$.
We neglect also  $\frac{d^2 \varphi}{dT^2} = \frac{d \Omega}{dT}$ compared to
$\frac{d \varphi}{dT} = { \Omega}$. This yields the simplified
expression
\be
\left ( \mathcal{F}^x_\mathbf{P} + i\, \mathcal{F}^y_\mathbf{P} \right )^{\rm leading\,\, order}
=-i\,  \frac{464}{105}\, f(\eta)\,
R^5\, \Omega^7\, e^{i\,\varphi}\,.
\label{P_flux_ip}
\ee
An important difference between  expression  (\ref{P_flux_ip}) and the
earlier quoted expressions for linear momentum fluxes, namely, Eqs.~(\ref{N_estimates}) and
(\ref{BQWflux}), is that 
the  proportionality to $ R^5\, \Omega^7$ was directly obtained from the 
original flux formula, Eq.~(\ref{P_flux_f}), {\em without explicitly using}
any ``Kepler-like equation'' linking $R$ to $\Omega$. We shall see later
that this difference significantly affects the estimate of the
final recoil velocity associated to the linear momentum
flux (\ref{P_flux_ip}).

  In order to obtain the velocity of the center of mass,
we then invoke linear momentum balance, namely,
\be
M\, \frac{ d }{dT} \left ( v^x_{\rm com} +i\, v^y_{\rm com} \right )
=  -\left ( \mathcal{F}^x_\mathbf{P} + i\,\mathcal{F}^y_\mathbf{P} \right )\,,
\label{com_b_f}
\ee
where $v^x_{\rm com} $ and $ v^y_{\rm com}$ are the $x$ and $y$ components of the center of mass
velocity vector $\mathbf{v}_{\rm com}$.

          At this stage, it is convenient to introduce rescaled, dimensionless 
radial, time and frequency variables, namely,
$ r = \frac{R}{M} $,  $t = \frac{T}{M}$, and $\omega = \frac{d\, \varphi}{dt} = M\, \Omega $.
This leads to the following differential equation for $ v^x_{\rm com} + i\, v^y_{\rm com}$
\be
\frac{ d }{dt} \left ( { v^x_{\rm com} +i\, v^y_{\rm com} }{} \right )
= i\, \frac{464}{105}\, f(\eta)\,
r^5\, \omega^7\, e^{i\,\varphi}\,.
\label{com_b_ip}
\ee
 
  This leading order ``quasi-Newtonian" result will be the basis of our investigation.
We shall also  discuss below how to use the 2PN correction terms derived
in Ref.~\cite{BQW05} [see Eq.~(\ref{BQWflux})]
 to improve the accuracy
of Eq.~(\ref{com_b_ip}).

  In this paper, in order to obtain an estimate for the velocity of the center of mass
during the late inspiral and subsequent plunge phases, we shall
numerically integrate  Eq.~(\ref{com_b_ip}) along with the differential equations that 
define the EOB dynamics.
 
  In the next section, we summarize the EOB dynamics applicable to non-spinning
compact binaries of arbitrary mass ratio moving in quasi-circular orbits during
the inspiral phase \citep{BD00}.
We also describe, in some detail, how we solve the relevant set of 
differential equations to obtain an EOB based estimate for the recoil during the late inspiral
and the subsequent plunge phases.
We shall also complement our numerical estimates by
 analytic arguments allowing one to understand in simple terms
 the main characteristics, and the order of magnitude, of our
 results.

\section{The late inspiral, plunge phases and the associated recoil using the EOB approach}
\label{S3}

 Let us first summarize the EOB approach relevant for describing the inspiral and plunge
phases of a compact binary. 
At the 2PN accuracy, the mapping between the full two-body
2PN dynamics, and the much simpler geodesic dynamics in the EOB metric,
given by Eq.~(\ref{geffective}), leads to an EOB dynamics described by the following Hamiltonian
(expressed in the scaled variables $ r \equiv \frac{R}{M}, t = \frac{T}{M},
\omega = \frac{d \varphi}{dt} = M\, \Omega$, and in polar coordinates)
\be
{\cal H} (r, p_r, p_{\varphi} ) = \frac{1}{\eta}\, \sqrt 
{ 1 + 2\, \eta \left [ \sqrt { A(r) \, \biggl ( 
1 + \frac{p_r^2}{B(r)} + \frac{ p_{\varphi}^2}{r^2} 
\biggr ) } -1 \right ] }\,,
\label{H_real}
\ee
where $A(r)$ and $B(r)$ [see Eqs.~(\ref{geffective}), and (\ref{eff_A_D}) above]
are given by
\bs
\bea
\label{A_r_d}
A(r) &=& 1 - \frac{2}{r} + \frac{2\, \eta}{r^3} \,,\\
B(r) &=& \frac{D(r)}{A(r)} = \frac{1}{A(r)} \left ( 1 - \frac{6\, \eta}{r^2} \right )\,.
\label{B_r_d}
\eea
\es
More precisely, the explicit form of the EOB equations of motion read
\bs
\bea
\label{eob_eq_rdot}
\frac{dr}{dt} &=& \frac{\partial {\mathcal H} (r, p_r, p_{\varphi} )
}{\partial p_r}\,,
\\
\label{eob_eq_omega}
\frac{d \varphi }{dt} &\equiv& \omega  = \frac{\partial \mathcal{H} (r, p_r, p_{\varphi} )
}{\partial p_{\varphi}}\,,\\
\label{eob_eq_prdot}
\frac{d p_r }{dt} &=& -\frac{\partial {\mathcal H} (r, p_r, p_{\varphi} )
}{\partial r }\,,\\
\label{eob_eq_jdot}
\frac{d p_{\varphi} }{dt} &=&  {\mathcal F}_{\varphi} \,.
\label{eob_eq_p_phi}
\eea
\label{eob_eq_s}
\es
The right-hand side of the last equation expresses the loss of angular momentum
under gravitational radiation reaction. Its explicit form will be discussed below.

  As mentioned earlier, to obtain an estimate for the recoil during the late inspiral and 
subsequent plunge, we solve along with the above set of differential equations, the one 
for $ v^x_{\rm com} + i\,v^y_{\rm com}$, given by Eq.~(\ref{com_b_ip}), namely,
\be
\label{Eq.19.old}
\frac{ d }{dt} \left ( { v^x_{\rm com} + i\, v^y_{\rm com} }{} \right )
= i\, \frac{464}{105}\, 
f(\eta)\, \, r^5\, \omega^7\, e^{i\,\varphi}\, \tilde F\,.
\ee
Here the supplementary factor $\tilde F = 1 + {\cal O} (v^2) + {\cal O} (v^3) +{\cal O} (v^4) $,
resulting from Ref.~\cite{BQW05}, is added to improve the accuracy of the leading-order,
quasi-Newtonian result, given by Eq.~(\ref{N_estimates}), to the 2PN level.
Its explicit form along  
our quasi-circular, sub LSO, orbits 
is  discussed in the following subsection.

\subsection{Inclusion of 2PN corrections in the fluxes of linear and angular momenta}

  In this subsection, we describe the construction of 2PN accurate expressions for
the recoil (linear momentum flux) factor $\tilde F$ in Eq.~(\ref{Eq.19.old}),
as well as for the 
radiation reaction force ${\mathcal F}_{\varphi}$ 
(the angular momentum flux), appearing in Eq.~(\ref{eob_eq_p_phi}).

Let us start by discussing the value of the correcting factor $\tilde F$
(and of its analog in the energy flux) during
 the adiabatic inspiral phase. During this phase, our construction
is facilitated by the fact that the orbital dynamics closely follows the
{\it one parameter sequence of stable circular orbits} that exists above the LSO.
In the EOB formalism, these orbits represent the {\it minima}, with respect to $r$, of
the Hamiltonian 
$ {\mathcal H}_{\rm circ}  (r, p_{\varphi} ) \equiv {\mathcal H} (r, p_r =0, p_{\varphi} )$.
Equivalently, we see from Eq.~(\ref{H_real}) that they are obtained by minimizing with respect to $r$
the effective potential 
\be
w ( r, p_{\varphi} ) = A(r) \left ( 1 + \frac{p_{\varphi}^2 }{r^2} \right )\,,
\label{weqn}
\ee
with $A(r)$ given by Eq.~(\ref{A_r_d}).
Minimizing  $w ( r, p_{\varphi} ) $ with respect to $r$ yields the following relation linking
$r$ to $ p_{\varphi}^2$ 
[see Eqs.~(4.5) and (4.6) of Ref.~\cite{BD00}]
\be
{p_{\varphi}^2 }|_{\rm circ} =  r \, \frac{ 1 - \frac{3\,\eta}{r^2}  } { 1 -  \frac{3}{r^2} 
+ \frac{5\,\eta}{r^3}\,.
} 
\label{pphicirc}
\ee
Inserting the latter result in the definition of the angular velocity, namely
$ \omega \equiv 
\frac{\partial \mathcal{H} (r, p_r, p_{\varphi} ) }{\partial p_{\varphi}} $, also 
considered along circular orbits [{i.e.}, 
$ \omega_{\rm circ}  \equiv
\frac{\partial \mathcal{H}_{\rm circ} (r,p_{\varphi} ) }{\partial p_{\varphi}} $],
then yields a relation connecting $ \omega_{\rm circ}$ to $r$. 
This  2PN generalization of Kepler's third law reads
[see Eqs. (4.8) of Ref.~\cite{BD00}]
\be
\omega_{\rm circ}^2 = \frac{1}{r^3} 
\biggl ( \frac{ 1 - \frac{3\,\eta }{r^2} }{ 1 + 2\, \eta\, \left ( \sqrt{w} -1 \right )}
\biggr )\,.
\label{kepler1}
\ee
In the test mass limit, $\eta \rightarrow 0$, we recover the well-known fact that 
circular orbits in a Schwarzschild geometry (in Schwarzschild coordinates)
satisfy the standard Kepler law: $ {\omega}^2\, r^3 =1$.
It is then traditional to use as PN order parameter 
$ v_{\omega} \equiv \omega^{1/3} = {\cal O} ( v/c) $, 
or equivalently $ x_{\omega} \equiv v_{\omega}^2 \equiv \omega^{2/3} = {\cal O} ( v^2/c^2) $,
to describe all possible PN corrections, be they proportional to the 
square of the linear azimuthal velocity $v_{\varphi} \equiv \omega\, r$, or to the 
gravitational potential $u \equiv \frac{1}{r} $. 
Indeed, when $\eta = 0$, we have the simple, Kepler-like links:
$ \left ( \omega\, r \right )^2 = \frac{1}{r} = v_{\omega}^2 = \omega^{2/3}$.
To extend these simple links to the comparable mass case $\eta \ne 0$ ( and to the sub-LSO 
quasi-circular orbits ), let us introduce the function
\be
{\psi} (r, p_{\varphi} ) \equiv \frac{ 1 + 2\, \eta \biggl (
 \sqrt{w (r, p_{\varphi}) } -1 \biggr )  } { 1- \frac{3\, \eta}{r^2} } \,,
\label{psieqn}
\ee
and the definition
\be
r_{\omega} \equiv r\, \biggl ( \psi ( r, p_{\varphi} ) \biggr )^{1/3}\,.
\label{romega}
\ee
These definitions are such that, along circular orbits, we can still write a simple Kepler-looking
law
\be
\omega^2\, r_{\omega}^3 = 1\,,
\label{kepler2}
\ee
as well as its usual consequences, such as 
$ \left ( \omega\, r_{\omega} \right )^2 = \frac{1}{ r_{\omega} } = \omega^{2/3} \equiv 
v_{\omega}^2 $.
We can then use these relations to rewrite any 2PN-accurate result expressed
(along circular orbits) in terms of $v_{\omega} \equiv \omega^{1/3} $ in terms 
of $\omega, r $ and $\psi$.

For instance, the 2PN accurate linear momentum flux \citep{BQW05},
 Eq.~(\ref{BQWflux}),  is proportional
to $ v_{\omega}^{11}\, F ( v_{\omega}) = \omega^7\, r_{\omega}^5\, F( \omega\, r_{\omega} )
= \omega^7\, \left ( r\, \psi^{1/3} \right )^5 F ( \omega\, r\, \psi^{1/3} ) $.
Our approach leads us  to considering that the basic
``quasi-Newtonian'' expression for the linear momentum flux is 
proportional to $\omega^7\, r^5$ [see Eq.~(\ref{P_flux_ip}) above]. In
other words, we are naturally led to writing the 2PN-accurate
flux in the form of Eq.~(\ref{Eq.19.old}) with a 2PN-correction
factor $\tilde F$  given by
\be
\tilde F (r,p_{\varphi}) = \biggl ( \psi(r,p_{\varphi}) \biggr )^{5/3}\, F ( \omega\, r\,
\psi^{1/3} )\,.
\label{Ftilde}
\ee
 Let us first note that for circular orbits {\it above the LSO} (for which all the above reasonings
are fully justified) the ``correcting factors'' linked to the function $\psi$ are very close to 1.
More precisely, if we consider the case $\eta = 0.2$ (which is the
 most important one)
  $\psi$ tends to $1$ when $ r \rightarrow \infty $, and as $r$ decreases
$\psi$ first decreases to reach a minimum $\psi_{\rm min} \simeq 0.9882$ around $r \simeq 9.2$.
Afterwards, it increases to reach $\psi_{\rm LSO} \simeq 0.9921$ when $ r = r_{\rm LSO} \simeq 5.8$.
Note that the factor  $\psi_{\rm LSO}^{1/3} \simeq 0.9974 $ modifying the azimuthal velocity
$v_{\varphi} = \omega\, r $ in Eq.~(\ref{Ftilde}) differs only by $\sim 3 \times 10^{-3} $
from {\it unity}. As for the total 2PN correcting factor $\tilde F$,
one can  see that it represents, above the LSO, a relatively modest modification of the quasi-Newtonian momentum flux. If we evaluate it
by inserting the straightforward 2PN-expanded version of the function
$F$, Eq.~(\ref{FTaylor}), into Eq.~(\ref{Ftilde}), we find a
result of order $1.24$ at the LSO. 

  Note also that, in the EOB approach, it is natural to consider as basic PN-ordering parameter
the azimuthal velocity,
\be
v_{\varphi} = \omega\, r \,,
\label{vphi}
\ee
which is an {\it invariantly defined} quantity \footnote{Indeed, 
in the EOB formalism, both $\omega$ and the effective-metric 
Schwarzschild radius $r$ are  invariant quantities.}.

  Up to this stage we have been assuming that we were considering
  quasi-circular orbits corresponding to a local minimum of the effective
radial potential. This happens when one is above the LSO. In contrast,
when considering the continuation of the orbit below the LSO, the
circular orbits [and their consequences, such as Eq.~(\ref{kepler1})]
are no longer physically relevant because they
correspond to unstable maxima of $ w(r, p_{\varphi})$, given by Eq.~(\ref{weqn}).
When considering quasi-circular orbits {\it below the LSO}, one should, in principle,
re-derive from scratch the 2PN-accurate linear momentum flux, {\it without  assuming}
any Kepler-like law of type Eq.~(\ref{kepler1}) or Eq.~(\ref{kepler2}).
The 2PN corrections then become functions of three independent variables:
$r, p_{\varphi},$ and $p_r$, or equivalently, $r, \omega,$ and $p_r$.
As pointed out in Refs.~\cite{BD99,BD00} the motion remains ``quasi-circular'' during the plunge
in the sense that the contributions linked to $p_r^2$ stay numerically small compared to 
those linked to $p_{\varphi}^2$, allowing one to neglect $p_r^2$.
The 2PN corrections during the plunge then become functions of two independent variables,
namely,
$r$ and $ p_{\varphi}$. In view of the arguments recalled above
pointing to a remarkable closeness between exact Einsteinian results
and quasi-Newtonian results expressed in terms of Schwarzschild-type
coordinates, we consider it likely that the momentum flux
during the plunge follows more or less the quasi-Newtonian
behavior ${\cal F} \propto r^5 \omega^7$. To ensure continuity with
the 2PN-correcting factor Eq.~(\ref{Ftilde}), which is present above
the LSO, we shall assume here that 2PN corrections below the LSO are sufficiently
well estimated by continuing to use the expression
 (\ref{Ftilde})\footnote{Note, however, that we no longer
assume the link, provided by Eq.~(\ref{pphicirc}). Indeed, this link exhibits an infinite growth
of $p_{\varphi}$ as $ r$ tends to the ($\eta$-modified ) ``light-ring'', where
$ r^3 - 3\, r^2 + 5\, \eta =0$. As we know instead that $p_{\varphi}$ stays below its LSO value during
the plunge (and evolves much more slowly than $r$), it seems a priori better to express
the PN corrections only in terms of well behaved quantities, such as $r, \omega,$ and $p_{\varphi}$.}.

  Note that this assumption differs from the one made in
Refs.~\cite{BD00} and \cite{BQW05}
which consisted in continuing to use the expressions giving the
2PN corrections as functions of $v_{\omega} \equiv \omega^{1/3}$.
Within the spirit of the EOB formalism, we feel that it is not very plausible to continue to 
use $v_{\omega}$ as basic PN ordering parameter,
and to express quantities only in terms of it. Indeed, the definition of
$v_{\omega}$ makes sense only so far as a Kepler-like law relating
$\omega$ to $r$ continues to hold. This is no longer the case
below the LSO (as will be illustrated in Fig. 2 below).
In absence of such a Kepler law, we prefer to remain close
to what is suggested by the leading-order quasi-Newtonian result
${\cal F} \propto r^5 \omega^7$. We shall further discuss below the
importance of this choice.

  Let us apply the same philosophy to the estimate of the radiation reaction term
${\cal F}_{\varphi}$, appearing in Eq.~(\ref{eob_eq_p_phi}),
below the LSO. It was shown in Refs.~\citep{DIS98} and \citep{BD00} that a
good estimate for ${\cal F}_{\varphi}$ {\it above the LSO} is given by
\be
{\cal F}^{\rm circ}_{\varphi} = - \frac{ 32}{5}\, \eta \, v_{\omega}^7\, \frac{ { \hat f}_{\rm DIS} (
v_{\omega}; \eta ) }{ 1 - \frac{v_{\omega} }{ v_{\rm pole} } }\,,
\label{Fphicirc}
\ee
where ${ \hat f}_{\rm DIS}$ is a 2.5PN ($v^5$-accurate) Pad\'e approximant for the angular momentum flux. It is defined,  {\it e.g},
in Eqs.~(3.28)-(3.36) of Ref.~\cite{BD00} [Note, however that the
2.5PN coefficient there must be corrected to a new value due to
Ref.~\cite{B05}.]. 
The definition of  ${ \hat f}_{\rm DIS}$  depends on the choice for the location of
the ``pole'' $v_{\rm pole}$. Following Ref.~\cite{DIS98} [and also Ref.~\cite{BD00}],
we use the value of
$v_{\rm pole}$ defined in Eq.~(3.37) of Ref.~\cite{BD00}.
[As shown in Ref.~\cite{DIS99} the precise choice of $v_{\rm pole}$ is not
very important and, for instance,
the value $v_{\rm pole} = v^{\rm \eta =0}_{\rm pole} = \frac{1}{\sqrt{3}} $ would suffice.]
Ref.~\cite{BD00} proposed to continue using Eq.~(\ref{Fphicirc}), expressed in terms of 
$v_{\omega} \equiv \omega^{1/3} $, even below the LSO. Here,  consistently with the arguments presented 
above, we shall instead use a different continuation
for ${\cal F}_{\varphi}$ below the LSO.
To derive it we need to know what is the analog, for the angular momentum loss, for the
``quasi-Newtonian'' result, displayed in Eq.~(\ref{P_flux_f}).
Consistently with what was  briefly mentioned above about the
energy flux, we know [see, e.g., Eq.~(4.23) of Ref.~\cite{KT80}] that the leading term in the angular momentum
loss $ \frac{d J_{\rm z}}{dt} $ is
$ \propto \,\, \stackrel{\!\!\!\!(2)}{I^{2-2} } \times \stackrel{\!\!\!\!(3)}{I^{22} }$. 
Remembering the leading-order expression for the quadrupole moment 
${I^{22} }$,
given in Eqs.~(\ref{is_s_ip}), we see that the ``quasi-Newtonian'' expression for ${\cal F}_{\varphi}$
is $ \propto
\frac{ d^2  \left ( R^2\, e^{2\,i\, \varphi} \right )}{ dT^2}  \times
\frac{ d^3  \left ( R^2\, e^{-2\,i\, \varphi} \right )}{ dT^3}  $ and therefore 
(in the quasi-circular approximation $\dot r \ll r\, \dot \varphi$)  $\propto r^4\, \omega^5$.
This shows that one should rewrite the leading factor
$v_{\omega}^7$ in Eq.~(\ref{Fphicirc}) as $\frac{v_{\omega}^5}{r_{\omega}} $, {i.e.},
$\frac{\left (  \omega\, r_{\omega} \right )^5}{r_{\omega}} =  \omega^5\, r_{\omega}^4 $.
In other words, this leads us to using, below the LSO, the following expression for 
the radiation reaction force ${\cal F}_{\varphi}$ in Eq.~(\ref{Fphicirc}),
\be
{\cal F}_{\varphi} (r, p_{\varphi} , p_r) =
 - \frac{ 32}{5}\, \eta \, {\omega}^5\,
\left ( r\, \psi^{1/3} \right ) ^4\,  
 \frac{ { \hat f}_{\rm DIS} 
( \omega\, r\, \psi^{1/3} ; \eta ) }{ 1 - \frac{ \omega\, r\, \psi^{1/3}  }{ v_{\rm pole} } }\,,
\label{radreac}
\ee
where the factor $\psi$ is a function of $r$ and $p_{\varphi}$, defined in Eq.~(\ref{psieqn})
and where $\omega = \omega(r, p_r,p_{\varphi})$ is defined by 
the second equation in Eqs.~(\ref{eob_eq_s}).

  Let us finally discuss the question of the re-summation of the 2PN-accurate 
correction factor, Eq.~(\ref{Ftilde}). When comparing the straightforward, PN-expanded versions
of the 2PN factor $F(v_{\omega})$, given by Eq.~(\ref{FTaylor}) and derived in
Ref.~\cite{BQW05},
entering the 2PN accurate linear momentum flux to its 
analog in the energy (or angular momentum) flux, namely,
\be
F_{\rm E}( v_{\omega}) = {\rm Taylor} \biggl [
\frac{ \hat f_{\rm DIS}(v_{\omega}) }{ 1 - \frac{v_{\omega}}{v_{\rm pole} } }
\biggr ]= 
1 + F_{\rm E2} ( \eta)\, v_{\omega}^2 +
 F_{\rm E3} ( \eta)\, v_{\omega}^3 +
 F_{\rm E4} ( \eta)\, v_{\omega}^4 +...\,,
\label{FEtaylor}
\ee
where $ F_{\rm E2} = -\left ( \frac{1247}{336} + \frac{35}{12}\, \eta \right )$,
$ F_{\rm E3} = 4\, \pi $, and $ F_{\rm E4} =
\left ( - \frac{44711}{9072} + \frac{9271}{504}\, \eta + \frac{65}{18}\, \eta^2 \right )
$ [see Ref.~\cite{PNfluxes}],
one notices that these two ``Taylor expansions'', {i.e.}, expressions for
$F(v_{\omega}) $ and $ F_{\rm E}( v_{\omega}) $, 
are rather similar. 
The corresponding Taylor coefficients $F_n$ [in $ \mathcal{F}^{i{({\rm BQW})} }_\mathbf{P}$]
and $F_{\rm En}$ [in ${\cal F}_{\rm E}$ or ${\cal F}_{\varphi }$] have the same signs, similar
sensitivities to the value of $\eta$, and roughly similar magnitudes.
Indeed, we can roughly consider that $ F_n \sim 1.3\, F_{{\rm E}n} $.
In addition, the same argument which was used in Ref.~\cite{DIS98} to show that the 
``exact'' function $F_{\rm E} (v)$ 
( analytically continued from its behavior above the LSO) has, in the limit $\eta \rightarrow 0$,
a pole at $v^{\eta=0}_{\rm pole} = \frac{1}{\sqrt{3}} $ ( {i.e.} at the light ring ) can
be applied to the function $F(v)$, appearing in 
Eq.~(\ref{BQWflux}) for $ \mathcal{F}^{i{({\rm BQW})} }_\mathbf{P}$,
to conclude that $F(v)$ also has, when $\eta \rightarrow 0$, a pole at the same 
(light-ring value) $v^{\eta=0}_{\rm pole} = \frac{1}{\sqrt{3}} $.
This suggests that a Pad\'e re-summation of the  $F(v)$, given by Eq.~(\ref{FTaylor}), might
improve the convergence behavior of $F(v)$, which is currently known only up to 2PN accuracy.
On the other hand,  the similarity between the two Taylor expansions $ F_n $ and $ F_{{\rm E}n}$
({i.e.} the fact that $ F_n \sim 1.3\, F_{{\rm E}n} $) suggests that 
both the successive Taylor and Pad\'e approximants of $F(v)$ will have convergence
properties similar to the corresponding approximants of $ F_{{\rm E}}$.
By looking at the convergence properties\footnote{Note that
the argument used in Ref.~\cite{BQW05} (namely:
the closeness of the $v^3$- and $v^4$-accurate Taylor approximants suggests that a good convergence
is reached with 2PN accuracy) is not conclusive in view of what happens for the similar
Taylor expansion of $F_E (v)$.
Indeed, one can easily check (say, when $\eta=0$ or $\eta =0.2$) that the $v^3$ 
and $v^4$ accurate Taylor approximants of $F_E(v)$ are close to each other, while
the next $v^5$-accurate, Taylor approximant of $F_E(v)$ is quite far away 
from both of them
(and also from the exact result, when $\eta \rightarrow 0$).
This is linked to the fact, emphasized in Ref.~\cite{DIS98}, that Taylor approximants
have rather erratic convergence properties as the PN order increases 
(while, by contrast, the Pad\'e approximants have a more monotonic
convergence, though they tend to accumulate somewhat below the exact result).
}
(when $\eta \rightarrow 0$) of the Taylor or Pad\'e
approximants of $ F_{{\rm E}}$, as displayed in Fig.~3 of Ref.~\cite{DIS98}, we observe that,
among the approximants of PN order $v^n$ with $ n \le 5$, the best one is the $v^5$-accurate
(2.5PN level) Pad\'e approximant
[{i.e} the one used in  Eq.~(\ref{Fphicirc}) above]. However,
the 2.5PN coefficient, $F_5 (\eta)$, is not currently known for the analogous linear-momentum flux
$ \mathcal{F}^{i{({\rm BQW})} }_\mathbf{P}$.  From Fig.~3 in Ref.~\cite{DIS98}, we expect the
$v^4$-accurate Taylor approximant of $F(v)$ to overestimate $F^{\rm exact} (v)$  
when $ v \le 0.4$ and to underestimate it when $v \ge 0.4$ 
(note that $v^{\rm LSO}_{\omega} ( \eta =0. 2) \simeq 0.414 $).
On the other hand, we expect the $v^4$-accurate Pad\'e approximant to $F(v)$ to follow 
$F^{\rm exact} (v)$ better, but to underestimate it for all values of $v$.
In the absence of knowledge about the 2.5PN contributions $F_5 (\eta)$, we shall compare here 
the results obtained both from using the $v^4$-accurate Taylor approximant, given by
Eq.~(\ref{FTaylor}), and its corresponding Pad\'e approximant, of the form
\be
F_{\rm P }(v) = \frac{ \hat g (v) } { 1- \frac{v}{v_{\rm pole} }} \,,
\ee
where we take, for simplicity, $v_{\rm pole} = \frac{1}{\sqrt{3}}$, and with
\be
\hat g (v) = \frac{1}{  1 +\frac{  c_1\,v}{  1 + \frac{ c_2\, v}
 {1 +
\frac{ c_3\, v}{
 1 + c_4\, v }
} } }\,.
\ee
Note that $\hat g (v) $ is constructed quite similarly to 
${\hat f}_{\rm DIS}(v) $,
{i.e.} by applying Eqs.~(3.29),(3.31),(3.34) and (3.35) of Ref.~\cite{BD00},
while replacing Eqs.~(3.32) and (3.33) in Ref.~\cite{BD00} by the coefficients
$F_2, F_3,$ and $F_4$ entering Eq.~(\ref{FTaylor}) above.

\subsection{Initial conditions for the dynamics}

  Before we present our numerical results, let us explain how we prescribe initial
conditions when solving these differential equations.
The initial conditions for $r$ and $\varphi$ are arbitrary and prescribe
the initial radial separation of the binary in the center of mass frame and its associated
angular position. The initial values for $p_r$ and $p_{\varphi}$ are obtained with the help
of the adiabatic approximation to the EOB inspiral, introduced in Sec.~IV (A) of Ref.~\cite{BD00}.
This approximation is obtained by imposing $p_r=0$ in the EOB dynamics, which implies that 
the effective body follows an adiabatic sequence of circular orbits 
with decreasing energy due to the emission of gravitational radiation.
This zeroth-order adiabatic approximation (which turns out to be enough
for our purpose) provides the following expressions for $p_{\varphi}$ and
$\omega $
\bs
\bea
p_{\varphi}|_{\rm adiab} &=& \biggl ( \frac{r^2\, ( r^2 - 3\, \eta)}{ r^3 -3\, r^2 + 5\, \eta} \biggr )^{1/2}\,,\\
\omega|_{\rm adiab} &=& 
\biggl ( \frac{ \left ( 1 - \frac{3\,\eta}{r^2} \right ) }{ r^3\, \left ( 1 + 2\,\eta \left [
\sqrt{z(r)} -1 \right ] \right ) } \biggr )^{1/2}\,,
\eea
\label{iv_eob}
\es
where $ z(r) = \frac{r^3\, A^2(r) }{ r^3 - 3\, r^2 + 5\, \eta} $ with $ A(r)$ given by
Eq.~(\ref{A_r_d}).

The initial value for $ v^x_{\rm com} + i\,v^y_{\rm com}$ is obtained by using 
\be
v^x_{\rm com} |_{\rm initial} 
= \frac{464}{105}\, \eta^2\, \sqrt{1-4\,\eta}\, \,\frac{\cos \varphi}{r^4}\,,
\,\,\,\,\,\,\,\,
v^y_{\rm com} |_{\rm initial} 
= \frac{464}{105}\, \eta^2\, \sqrt{1-4\,\eta}\, \,\frac{\sin \varphi}{r^4}\,,
\label{iv_v_com}
\ee
derivable from Eqs.~(\ref{N_estimates}).
In the next subsection, we present analytical insights into the
physical behavior underlying our numerical
estimates.

\subsection{Linear momentum loss during inspiral and plunge}

  Before numerically implementing our strategy for estimating the recoil during the late 
inspiral, the subsequent plunge and the final merger, let us outline
the main physical features of our calculation\footnote{See below for the effects of ring-down.}.
Let us first note that the final
recoil velocity is essentially given by an integral of the form
\be
\left (  v^x_{\rm com} + i\, v^y_{\rm com} \right )^{\rm terminal}
=  i \, {\cal I} \equiv
i \, \int^{+\infty}_{-\infty} dt\, a(t)\, e^{i\, \varphi(t)}\,,
\label{v_f_int}
\ee
where $\varphi(t)$ is the orbital phase, while the ``amplitude"
$a(t)= | \mathcal{F}_\mathbf{P}| $
is proportional to $r^5\, \omega^7\, \tilde F $. The first important point to realize
is that the value of the above integral ${\cal I}$ is dominated by what happens
in the time intervals where the amplitude $a(t)$ varies in a ``non-adiabatic" manner.
Indeed, let us first assume that the amplitude $a(t)$ always varies in an adiabatic manner
with respect the orbital phase $\varphi(t)$, {i.e} 
let
$ \frac{\dot a }{a} \ll \dot \varphi $.
This can be conveniently formalized by replacing the phase factor $e^{i\, \varphi} $
by $e^{i\, \varphi/ \epsilon} $, where $\epsilon$ is a formal ``small parameter" measuring
how small the ratio $ \frac{\dot a }{a} / \frac{\dot \varphi }{\epsilon} \sim {\cal O} (\epsilon)
$ is. Using the fact that $\varphi (t) $ is a monotonic function of time, one can use
$\varphi$, instead of $t$, in the above integral.
After replacing $a(t)$ by ${\cal A} (\varphi ) \equiv \frac{ a(t) }{ \dot \varphi (t)} $, we
get ${\cal I} (\epsilon) = \int^{+\infty}_{-\infty} d {\varphi}  {\cal A } (\varphi)\,
e^{i\, \varphi(t)/ \epsilon} $. Using
$e^{i\, \varphi/ \epsilon} =
\frac{\epsilon}{i}\,
\left ( \frac{d \,\,\,e^{i\, \varphi/\epsilon} } { d\, \varphi}  \right ) $
and integrating by parts, we find  $ {\cal I} (\epsilon) = i\, \epsilon \, \int d \varphi\,
{\cal A}' (\varphi ) \, e^{i\, \varphi/\epsilon} $, where we employed
the vanishing of $ {\cal A}(\varphi) $ at $\pm \infty $
(see below). Repeating this procedure, we get
\be
{\cal I} (\epsilon) = \left ( i\, \epsilon \right ) ^{n} \, \int^{+\infty}_{-\infty}
d \varphi\, {\cal A}^{(n)} (\varphi ) \, e^{i\, \varphi/\epsilon}\,,
\ee
where ${\cal A}^{(n)} (\varphi )$ is the $n$-th derivative of ${\cal A} $ w.r.t ${\varphi}$.
This (well-known) result means that, when $\epsilon \rightarrow 0$, $ {\cal I} (\epsilon) =
{\cal O} (\epsilon^n)$ for any integer $n$. In other words, ${\cal I}(\epsilon) $
{\em vanishes faster than any power of $\epsilon$}, if $a(t)$ varies adiabatically during the
whole process. In most cases, this means that ${\cal I} (\epsilon) $ is exponentially
small $\sim e^{ - C/\epsilon^2} $, and therefore numerically negligible compared
to the naive estimates of the type 
$ {\cal I} \sim a_{\rm max}\, \Delta t$, where $\Delta t $ is a characteristic variation time,
say $\Delta t \sim \frac{1}{\omega}$ with $\omega = \dot \varphi$,
 that one might have been
tempted to make.

 This mathematical reminder shows that the actual magnitude of the momentum flux
$| \mathcal{F}_\mathbf{P}| = a(t) $ during the inspiral and the plunge is
secondary with respect to the question of knowing the characteristic 
time-scale on which $|\mathcal{F}_\mathbf{P}|$ varies during the plunge. If
$\frac{\dot a}{a} $ stayed always small compared to the orbital 
frequency $\omega = \dot \varphi$, the recoil would be 
a non-pertubative effect; and it would be practically
hopeless to try to estimate it by starting from approximate analytical 
expressions. [On the other hand, we would know that the recoil is exponentially small, so that 
it would be astrophysically negligible.]
However, the study of the time evolution of the amplitude $a(t)$ during the EOB plunge
shows that, while it remains adiabatic ($ \frac{\dot a } {a} \ll \dot \varphi $)
during most of the inspiral and plunge, it becomes barely non-adiabatic near the
moment where $a(t)$ reaches a {\em maximum} [at which point, the criterion for 
non-adiabaticity must involve the {\em second} time derivative of $a(t)$].
This is illustrated in Fig.~\ref{Fig_1_MF_r_t},
which shows the evolution of
the magnitude of the momentum flux,  
$|\mathcal{F}_\mathbf{P}|$, during the late inspiral and the plunge.
This figure makes it clear that the characteristic evolution time scale for 
$a(t)$ is shortest near its maximum, {i.e.} for a (scaled) radius $r_{\rm max} \simeq
3.5 $. Before discussing the consequences of this fact, let us outline how one can
analytically understand why $|\mathcal{F}_\mathbf{P}|$ has the behavior exhibited
in Fig.~\ref{Fig_1_MF_r_t}.

  The main factor determining the behavior of $|\mathcal{F}_\mathbf{P}|$ is the product 
$ r^5\, \omega^7 $ in Eq.~(\ref{com_b_ip}). During the plunge, the evolution of $r$ and $ \omega$ are
governed by the EOB equations of motion, namely Eqs.~(\ref{eob_eq_s}). 
In these equations, the radiative damping ``force" ${\mathcal F}_{\varphi}$ is crucial
to drive the slow inspiral and to trigger the plunge, but was found to play a minor role 
once the plunge is well on its way \citep{BD00,BD_MG}. As a consequence, the evolution of
$\omega$ during the plunge is approximately given by $\omega = \frac{\partial {\mathcal H} (r, p_r, p_{\varphi} )}{ \partial p_{\varphi} } $, with $ p_{\varphi} \simeq $ constant as well as
$ {\mathcal H} \simeq $ constant  ( zero-damping approximation).
In this approximation, one thereby finds that $\omega$ is approximately
proportional to the ratio $ \frac{A(r)}{r^2} $ so that we can write,
during the plunge, the approximate link

\be\label{keplerplunge}
\omega_{\rm plunge}(r) \approx \omega_{\rm LSO}
\frac{r_{\rm LSO}^2}{A(r_{\rm LSO})} \frac{A(r)}{r^2} \,.
\ee
The change in behavior of $\omega(r)$ between the inspiral (where
Kepler's third law $\omega^2 r^3 \approx {\rm constant} $ holds approximately),
and the plunge [where Eq.~(\ref{keplerplunge}) holds instead]
is illustrated in Fig.~\ref{Fig_2}.

  As a consequence of Eq.~(\ref{keplerplunge}), we find that the
behavior of the linear momentum flux during the plunge is
approximately given by
\be \label{analyticFp}
|\mathcal{F}_\mathbf{P}|  \propto  r^5\, \omega^7 \,  \propto  
\frac{1}{r^9}\, \left ( 1 - \frac{2}{r} + \frac{2\, \eta}{r^3} \right )^7\,.
\ee
Denoting $u= \frac{1}{r} $, it is easily seen that $|\mathcal{F}_\mathbf{P}| \propto  
u^9\, \left ( 1 - 2\, u + 2\, \eta\, u^3 \right )^7 $
reaches a maximum value when  $\left (  9 - 32\, u_{\rm max}  + 60\, \eta\, u_{\rm max}^3  \right ) = 0$.
When $\eta =0.2$ ( which  corresponds to the maximum of the overall 
factor $f(\eta)$, given by Eq.~(\ref{feta}),
and thereby approximately to the maximum recoil ),
this gives $u_{\rm max} \simeq 0.29044$, corresponding to $r^{\rm analytic}_{\rm max} = \frac{1}{u_{\rm max} }
\simeq 3.4431$.
This analytical argument agrees well with our numerical results.
Indeed, we find that $|\mathcal{F}_\mathbf{P}|$, computed using
the quasi-Newtonian version of Eq.~(\ref{Eq.19.old}),
has a maximum around $r^{\rm numerical}_{ \rm max} \simeq  3.501 $. 
One can even go further and analytically
study the behavior of $|\mathcal{F}_\mathbf{P}|$ near its maximum.
The most important time scale there is defined by the curvature of
$|\mathcal{F}_\mathbf{P}|$ near its maximum:
\be
\tau_{\rm max}^2 \equiv - \frac{ \mathcal{F}^{\rm max} _\mathbf{P} }
{ \left ( \frac{ d^2 \mathcal{F}_\mathbf{P} }{ dt^2 } \right )^{\rm max} } \,,
\ee
where, for notational simplicity, we henceforth denote by $\mathcal{F}_\mathbf{P}$
the {\em modulus} of the momentum flux.

An important dimensionless quantity associated to the time
scale $\tau_{\rm max}$ is the ``quality factor" $Q \equiv \omega_{\rm max} \, \tau_{\rm max} $
associated to the ``resonance peak" of $ \mathcal{F}_\mathbf{P}$.
Indeed, values of $Q$ of order unity mean (as we shall find)
that the evolution of $\mathcal{F}_\mathbf{P}$
near its maximum is just fast enough to be non-adiabatic there.


 In view of the discussion above,
this means that the recoil, {i.e.} (modulo a factor $i$)
the integral ${\cal I}$, given by Eq.~(\ref{v_f_int}),
will be dominated by what happens near the maximum of $ \mathcal{F}_\mathbf{P}$.
Therefore, we can analytically estimate ${\cal I}$ by replacing $ a(t) =
\mathcal{F}_\mathbf{P} ( \equiv | \mathcal{F}_\mathbf{P}|)$
by the local approximation
( with $\bar t \equiv t - t_{\rm max} $ )
\be
a(t) \simeq \mathcal{F}^{\rm max}_\mathbf{P} + 
\frac{1}{2}\, \left ( \frac{ d^2 \mathcal{F}_\mathbf{P} }{ dt^2 } \right )^{\rm max} \, \bar t^2 
= \mathcal{F}^{\rm max}_\mathbf{P} \, \left ( 1 - \frac{1}{2}\, \frac{ \bar t^2 }{ \tau_{\rm max}^2 }
\right ) 
\simeq \mathcal{F}^{\rm max}_\mathbf{P}\, e^{- \frac{ \bar t^2 }{ 2\, \tau_{\rm max}^2 }}\,,
\ee
and the phase $\varphi (t)$ by
\be
\varphi (t) \simeq \varphi_{\rm max} + \dot \varphi_{\rm max}\, \bar t 
+ \frac{1}{2}\, \ddot \varphi_{\rm max}\, \bar t^2 = \varphi_{\rm max}
+ \omega_{\rm max} \, \bar t + \frac{1}{2}\, \dot \omega_{\rm max} \, \bar t^2 \,.
\ee
If we then consider the recoil acquired up to some given time $\bar t'$, its is given by a truncated
Gaussian integral
\be
\left ( v^x_{\rm com} +i\, v^y_{\rm com} \right ) ( \bar t') \simeq i \, \mathcal{F}^{\rm max}_\mathbf{P}\, e^{i\, \varphi_{\rm max}}
\int_{-\infty}^{\bar t'} d \bar t \, e^{ -\frac{ 1}{2}\, \alpha\, \bar t^2  + \beta\, \bar t } \,,
\label{tr_GI}
\ee
where (posing $\epsilon_{\rm max } \equiv \dot \omega_{\rm max}\, \tau^2_{\rm max} $ )
\be
\alpha =  \frac{1}{\tau^2_{\rm max} } - i\, \dot \omega_{\rm max} \equiv
\frac{1}{\tau^2_{\rm max} } \left ( 1 - i\, \epsilon_{\rm max } \right )\,,
\,\,\, \,\,\,
\beta = i\, \omega_{\rm max}\,.
\label{XY_def}
\ee
When $\bar t'$ gets positive and large with respect  to $\tau_{\rm max}$
( so that $\bar t' $ is effectively $\sim + \infty $ ), we can estimate the total
integrated recoil using the standard complex Gaussian integral:
\be
\int_{-\infty}^{+ \infty}
dy e^{-\frac{1}{2} \, {\alpha} \, y^2 + \beta\, y } = \sqrt{2\, \pi} \,\,
\frac{ e^{\frac{1}{2}\, \frac{\beta^2}{ \alpha} } }{ \sqrt{\alpha}}\,.
\ee
This yields, for the modulus of the corresponding total integrated recoil
\be
| v^x_{\rm com} +i\, v^y_{\rm com} | \simeq \sqrt{2\, \pi} \,\,
\mathcal{F}^{\rm max}_\mathbf{P}  \,\, \frac{ \tau_{\rm max} } 
{ \left ( 1 + \epsilon^2_{\rm max} 
\right )^{1/4}  } \,\,  e ^{ - \frac{1}{2}\,
\frac{ \omega^2_{\rm max}\, \tau^2_{\rm max} }{ 1 + \epsilon^2_{\rm max} } } \,.
\label{v_com_estimate}
\ee
This approximate analytical result vividly illustrates the preceding discussion. Indeed,
if the evolution of $ \mathcal{F}_\mathbf{P} (t) $
were adiabatic ($ Q \equiv \omega_{\rm max}\, \tau_{\rm max} \gg 1 $ ) the total integrated recoil
would be exponentially small ( even if $\mathcal{F}^{\rm max}_\mathbf{P}$ gets large ).

We have already indicated above, see Eq.~(\ref{analyticFp}),
how one can analytically determine the location on the $r$-axis
of the maximum of $ \mathcal{F}_\mathbf{P}$. By analytically
expanding Eq.~(\ref{analyticFp}) around its maximum, one can also
get an analytical expression for the product
$\tau_{\rm max}\,{\dot r}_{\rm max}$. As the reasoning above
also gave the variation with $r$ of the angular frequency,
namely, $\omega(r)/ \omega_{LSO} \approx [A(r)/r^2]/[A(r_{LSO})/r_{LSO}^2]$,
we can obtain analytical estimates of $\omega_{\rm max}$ and of
$ \dot {\omega}_{\rm max}/ {\dot r}_{\rm max}$. Finally, to get
analytical estimates of all the quantities entering
Eq.~(\ref{v_com_estimate}) we need an analytical estimate of
$\dot r$ around $ r=r_{\rm max}$. This can be obtained (though
only with modest accuracy) by
using again the zero-damping approximation to write that the
energy is approximately conserved during the plunge:
hence
\be \label{prplunge}
 A(r) \biggl [  1 + p_r^2/B(r) + p_{\varphi}^2/r^2 \biggr ] \approx w_{LSO} \,,
\ee
with $w_{LSO} \approx A(r_{LSO}) ( 1 +p_{\varphi LSO}^2/r_{LSO}^2)$.
Eq.~(\ref{prplunge}) approximately determines the value of
$p_r$ during the plunge. From it, one then deduces the value of $\dot r$
by using the Hamilton equation, Eq.~(\ref{eob_eq_rdot}). 
See Fig.~4 of Ref.~\cite{BD00} for a plot of ${\dot r} $ during the plunge.

These analytical approximations allows one to obtain estimates
for all the quantities entering the crucial Eq.~(\ref{v_com_estimate}),
and thereby to obtain an analytical estimate of the expected total
recoil velocity $v_{\rm com}$. We found that the results agrees
within a few percent with the numbers one can extract from our full
numerical simulations. The complete set of relevant quantities,
extracted from our simulations, for the behavior around
the time where $| \mathcal{F}_\mathbf{P} | $ reaches its maximum value
 are (for $\eta = 0.2$):
 $  r_{\rm max} \simeq 3.501, {\dot r}_{\rm max} \simeq -0.113, 
 \omega_{\rm max} \simeq 0.1255 ,
{\dot \omega}_{\rm max} \simeq 2.952 \times 10^{-3},
| \mathcal{F}_\mathbf{P} |_{\rm max} \simeq 2.039 \times 10^{-5},
| \ddot {\mathcal{F}_\mathbf{P}} |_{\rm max} \simeq 3.683  \times 10^{-7},
\tau \simeq 7.440, Q \simeq 0.9337 , \epsilon \simeq 0.1634,$ and
$ \psi \simeq 1.007   $.
Of particular importance is the
value of the quality factor, namely, $Q \simeq 0.9337$. The fact that it is
of order unity means that a net integrated recoil is acquired soon
after $ \mathcal{F}_\mathbf{P} (t) $ reaches its maximum value,
i.e. soon after the plunge has fallen below $r \simeq 3.501$, and
therefore {\em before} reaching the light ring radius $r_{\rm lr} = 3$.

   Finally,
we can analytically estimate, using Eq.~(\ref{v_com_estimate}), the final recoil that one might expect.
We find
\be
|  v^x_{\rm com} +i\, v^y_{\rm com} |^{\rm integrated} \simeq
74.06 \, \frac{ f (\eta) }{ f(0.2)} \, \tilde F_{\rm max}  \, {\rm km/s} \,,
\label{ana_est_v_com}
\ee
where we recall that $ f(0.2) = 0.0178885 $,
the maximum value reached by $ f(\eta)$ (when $\eta =0.2$),
and where $ \tilde F_{\rm max} $ denotes the value of
the 2PN correction factor at $ r = r_{\rm max} \simeq 3.501$.
By definition, the quasi-Newtonian estimate corresponds to
taking $\tilde F \equiv 1 $.

\section{Transition from plunge to ring-down, and gravitational recoil during ring-down}
\label{S4}

  Though the analytical estimate, given by Eq.~(\ref{v_com_estimate}), is interesting by the physical 
information it conveys [effect dominated by the maximum of 
$  \mathcal{F}_\mathbf{P} (t) $, dependence on $f(\eta)$
and $\tilde F_{\rm max}$, 
and the obtainment of a small pure number from high powers of numbers
``of order unity''], let us hasten to add that it is only an approximation to the
real terminal recoil. Indeed, the above estimate, 
was obtained by taking the formal limit 
$\bar t \rightarrow + \infty $ in the truncated Gaussian integral, Eq.~(\ref{tr_GI}). 
However, as we have already indicated in the introduction, the physics behind the
approximate analytical formulas,
Eqs.~(\ref{P_flux_ip}), (\ref{com_b_ip}), or
(\ref{Eq.19.old}),
changes when $r$ reaches 
the ``light ring'' $r \simeq 3$.
Following the analogous estimate of complete waveforms in Ref.~\cite{BD00}, we propose here to estimate
the contribution to the recoil due to the merger of black holes by formally terminating the plunge
when the scaled radial coordinate gets around $r \simeq 3$, and by matching 
there the relevant time derivatives of the radiative multipole moments during 
the late plunge phase to corresponding ``ring-down'' multipole moments, constructed from 
appropriate quasi-normal mode contributions.

  Let us first discuss why it is important to ensure as smooth a matching as possible during 
the transition from plunge to ring-down. To see this, let us consider again the 
approximate form of the final recoil velocity, given by Eq.~(\ref{v_f_int}),
but let us now divide the full time interval in two phases:
an inspiral + plunge phase, lasting from $t = -\infty$ up to some $t_{\rm match}$,
followed by a ring-down phase, lasting from $t_{\rm match}$ up to $t = +\infty$.
The total  recoil will be the sum of two contributions of the form
\footnote{Actually, the integral during the ring-down phase is a sum of terms 
$\propto e^{+\,i\, \varphi} $ and $ e^{-\,i\, \varphi} $. See discussion below.}
\be
{\cal I}_{\rm plunge} = 
\int^{t_{\rm match} }_{-\infty} dt\, 
\left ( 
a(t) \, e^{i\, \varphi(t) } \right )|_{\rm plunge} \,;\,\,\,
{\cal I}_{\rm ring} = 
 \int^{+\infty}_{t_{\rm match}  } dt\, 
\left ( 
a(t) \, e^{i\, \varphi(t) } \right )|_{\rm ring}\,.
\label{edgeintegral}
\ee
We now focus on the contribution to ${\cal I}_{\rm tot} = {\cal I}_{\rm plunge} + {\cal I}_{\rm ring}$
that is formally linked to any ``mismatch'' between the two behaviors of the linear momentum flux
around $t= t_{\rm match}$, {i.e.} to any discontinuity between
 $i \, \left ( \mathcal{F}^{x \rm plunge}_\mathbf{P}
 + i \, \mathcal{F}^{y \rm plunge}_\mathbf{P} \right )= \left [
a(t) \, e^{i\, \varphi(t) } \right ]_{\rm plunge} $, considered for $t < t_{\rm match}$,
and
$i \, \left ( \mathcal{F}^{x \rm ring}_\mathbf{P}
 + i \, \mathcal{F}^{y \rm ring}_\mathbf{P} \right ) = \left [
a(t) \, e^{i\, \varphi(t) } \right ]_{\rm ring} $, considered for $t > t_{\rm match}$.
The effect of any discontinuity around $t= t_{\rm match} $
can be obtained by summing the ``edge contributions'' of the two semi-infinite 
integrals, Eq.~(\ref{edgeintegral}), {i.e.} the contributions linked to the upper or lower 
cut-off $t = t_{\rm match}$.
These ``edge contributions'' have been worked out in 
Ref.~\cite{DIS99} to next-to-leading order
in ``adiabaticity expansion'' [{i.e.} in powers of the formal small parameter $\epsilon$,
introduced in Eq.~(\ref{v_f_int}) above, by replacing $e^{i\,\varphi}$ by $e^{i\,\varphi/\epsilon}$ ], by using the ``integration by parts'' technique introduced above for
showing that, in the absence of any discontinuity, the integral
${\cal I}(\epsilon) = \int dt\, a(t)\, e^{i\,\varphi(t)/\epsilon} $  vanishes faster than any power of 
$\epsilon$. Adding two terms of the type of Eq.~(3.17) in Ref.~\cite{DIS99}, the total edge 
contribution is of the form
\be
{\cal I}^{\rm edge}_{\rm plunge} + {\cal I}^{\rm edge}_{\rm ring}  
= \biggl [  \frac{a(t)}{ i\, \dot \varphi (t)}\, e^{i\, \varphi(t)}
\biggl \{ 
1 + \frac{1}{i\, \dot \varphi (t)}\, e^{i\, \varphi(t)}
\biggl [  \frac{\ddot \varphi (t) }{ \dot \varphi(t)}
-  \frac{\dot a(t) }{ a(t) }
\biggr ]
\biggr \}
\biggr ]^{\rm plunge}_{\rm ring}\,,
\label{i_edge_p_r}
\ee

where the square bracket $\biggl [  {\cal F}( a(t),
\varphi (t),
\dot \varphi (t),...)\biggr ]^{\rm plunge}_{\rm ring}$
on the right-hand side of the above equation denotes
the {\it difference},
$ {\cal F}( a_{\rm plunge}(t_{\rm match} ),
\varphi_{\rm plunge} (t_{\rm match} ),
\dot \varphi_{\rm plunge} (t_{\rm match}),...) - $
$ {\cal F}( a_{\rm ring}(t_{\rm match} ),
\varphi_{\rm ring} (t_{\rm match} ),
\dot \varphi_{\rm ring} (t_{\rm match}),...) $.
This analytical result highlights the following fact: any discontinuity between the
amplitude, the phase, or any of their time-derivatives
across $t_{\rm match}$ will contribute to the final recoil velocity.
Therefore, if we want to minimize the {\it spurious} effects linked to our 
describing the smooth transition
between the plunge and the merger by a fictitious sharp transition happening at $t = t_{\rm match}$,
we should try to match as many derivatives as possible of 
$\mathcal{F}_\mathbf{P} \propto a(t)\, e^{i\, \varphi(t)} $ across $t = t_{\rm match}$.
On the other hand, we are going to see that, even after having matched as well as possible
$\mathcal{F}_\mathbf{P}$ across $t = t_{\rm match}$, there remains a (non-spurious) 
``edge'' contribution linked to the physical change of behavior across $t = t_{\rm match}$.
 
   To see this, let us consider in more detail how one can implement a physically motivated matching 
across $t = t_{\rm match}$ ( corresponding to $r_{\rm match} \simeq 3$). All the physical 
effects which are important for the present study ( flux of energy related to 
$\mathcal{F}_{\varphi}$, and flux of linear momentum, $\mathcal{F}_\mathbf{P} $)
can be expressed as integrals over a sphere at infinity with integrands proportional
to the local gravitational wave energy flux $ \frac{dE}{d\Omega\,dt} = r^2\, T^{\rm GW}_{00}
\propto \left ( r\, \dot h_{ij} \right )^2 $, where $h_{ij}$ is the TT-gauge dimensionless gravitational
wave amplitude, and $\dot h_{ij}$ is its time derivative $\frac{\partial h_{ij} }{\partial t}$ 
[see, for {e.g.}, Ref.~\cite{KT80}]. This motivates us to try to match as well as possible
the quantity $r\, \dot h_{ij}(t,r,\theta, \phi)$, where 
$\theta$ and $\phi$ are polar angles on the sphere at infinity, between plunge and ring-down.
The ``radiative multipole moments'' that enter the multipole expansion of $r\, \dot h_{ij}$ are,
by definition, the $(l+1)$-{th} time derivatives of the $l$-th mass ($I^{lm}$) and spin
(or current) ($S^{lm}$) multipole moments
\footnote{For simplicity, we use here the nomenclature of Ref.~\cite{KT80}.
In the multipolar post-Minkowskian formalism, Refs.~\cite{
BD86,BD89,DI91,BD92,LB95,LB98,BIJ2002,BDEI04,ABIQ04}, the ``radiative'' moments are 
defined as $U^{lm} \sim   \stackrel{\!\!\!\!(l)}{I^{lm} } $ and 
$V^{lm} \sim   \stackrel{\!\!\!\!(l)}{S^{lm} } $, { i.e.} as the moments entering the multipole
expansion  of $r\, h_{ij}$. In the latter nomenclature, the moments that most directly
enter the quantities that we need would be $\dot U^{lm}$ and $\dot V^{lm}$
(and would include all required `tail' effects).
}.  It is therefore most natural to match  $\stackrel{\!\!\!\!(l+1)}{I^{lm} }$
and $\stackrel{\!\!\!\!(l+1)}{S^{lm} }$ across $t = t_{\rm match}$.
For the evaluation of $\mathcal{F}_\mathbf{P}$ at the leading order, 
the relevant radiative moments, as seen in Eq.~(\ref{P_flux_f}), are
$ \stackrel{\!\!\!\!(3)}{I^{2 \pm 2} }; 
\stackrel{\!\!\!\!(3)}{S^{2 \pm 1} }; \stackrel{\!\!\!\!(4)}{I^{3 \pm 1} }$ 
and $\stackrel{\!\!\!\!(4)}{I^{3 \pm 3} }$.
These terms correspond to gravitational waves, emitted by the two black holes, of multipolarity:
( $l=2$, $m = \pm 2$, even parity); ( $l=2$, $m = \pm 1$, odd parity); 
( $l=3$, $m = \pm 1$, even parity); ( $l=3$, $m = \pm 3$, even parity),
respectively. As a first approximation
\footnote{We leave to future work the refinement consisting in using modes
propagating over a Kerr background.}, we can consider that these gravitational waves
propagate (for radii larger than the radial distance $r(t)$ separating the two black holes)
on a Schwarzschild background, of mass $M_{\rm S} = E^{\rm real}_{\rm tot} \sim M \equiv m_1 +m_2$,
approximately representing the (physical) spacetime outside the two holes.
Therefore, when $r(t)$ gets smaller than about $3$, the relevant modes $\dot h^{l,m}_{ij} $
will be strongly filtered by the corresponding Regge-Wheeler-Zerilli effective 
potential $V^{(\rm even/odd)}_{lm}(r) $.  This filtering can be approximated by saying that,
when the source of a mode $(l,m, \pi)$, where $\pi$ denotes the parity,
falls below $r=3$, the corresponding outgoing wave mode can be described by a 
superposition of quasi-normal modes (QNM's) of the same multipolarity $(l,m, \pi)$.

Several nice simplifying features of gravitational wave propagation on a Schwarzschild background
are that: (i) the effective potential $V^{\pi}_{l}(r)$ does not depend on the 
``magnetic quantum number'', $m$, (ii) $V^{\pi}_{l}(r)$ is real, and (iii) though
$V^{\rm even}_l \ne V^{\rm odd}_l$, they have the same spectrum of QNM complex frequencies
[for a review of QNM's see Ref.~\cite{KS_LR}].
For each value of the multipolar order $l$, there is a double infinite 
sequence of QNM complex frequencies, say
\be
\sigma^{\pm}_{ln} = \alpha_{ln} \pm i\, \omega_{ln} \,,
\label{sigmaeqn}
\ee
where $n =0,1,2,...$,and $\alpha_{ln} $ and  $\omega_{ln} $ are both real and positive [so that
$\sigma^{+}_{ln} \equiv \left ( \sigma^{-}_{ln} \right )^{*} $].
The notation here is that the $n$-{th} QNM mode belonging to the multipolarity $l$ decays, when
$ t \rightarrow + \infty $, proportionally to 
$ e^{-\sigma^{\pm}_{ln}\, t} = e^{-\alpha_{ln}\, t} \, e^{ \mp i \omega_{ln}\, t} $.
For each value of $l$ the fundamental QNM mode $n=0$ is the least-damped one, { i.e.} the one with the
smallest value for $\alpha_{ln}$.

  Finally, our matching procedure consists in joining, as smoothly as possible, 
across $t = t_{\rm match}$,  each relevant multipolar mode entering
$r\,\dot h^{l,m}_{ij} $, namely, 
$ \stackrel{\!\!\!\!(3)}{I^{2 \pm 2}_{\rm plunge}(t) },
\stackrel{\!\!\!\!(3)}{S^{2 \pm 1}_{\rm plunge}(t) }, \stackrel{\!\!\!\!(4)}{I^{3 \pm 1}_{\rm plunge}(t)}$,
and $\stackrel{\!\!\!\!(4)}{I^{3 \pm 3}_{\rm plunge}(t) }$, obtained for $t < t_{\rm match}$
by differentiating Eqs.~(\ref{is_s_ip}) in the quasi-circular approximation
($ \dot r \ll r\, \dot \phi$), to corresponding ``ring down'' multipole moments,
made of sum of decaying QNM modes.
For instance, this leads (after scaling out the total mass $M$) to matching
\be
\stackrel{\!\!\!\!(3)}{I^{22}_{\rm plunge} }(t) = i\, \frac{16}{5}\, \sqrt{10\,\pi}\,\, \eta\, r(t)^2\,
\omega(t)^3\, e^{-2\,i\,\varphi(t)}\,\,\,\,\,\, ({\rm for  } \,\,\,t < t_{\rm match} )\,,
\label{i_3_22_S}
\ee
where $ r(t), \varphi(t), \omega(t) = \dot \varphi(t)$ are obtained by numerically
integrating the EOB dynamics, Eqs.~(\ref{eob_eq_s}), 
to a corresponding ``ring down'' radiative moment of the 
form 
\be
\stackrel{\!\!\!\!(3)}{I^{22}_{\rm ring} } (t) =
\sum_{n =0,1,..} \left \{ C^{+}_{n} ( I^{22} )\, e^{- \sigma^{+}_{2n}\, \tau }
+ C^{-}_{n} ( I^{22} )\, e^{- \sigma^{-}_{2n}\, \tau} \right \}
\,\,\,\,\, ( {\rm for  } \,\,\,\tau \equiv t-t_{\rm match} >0  )\,,
\label{i_3_22_rd}
\ee
where $\sigma^{\pm}_{2n} $, $n=0,1,..$, are the QNM frequencies, Eq.~(\ref{sigmaeqn}), 
belonging to the multipolarity $l=2$,
and where $C^{\pm}_{n}(I^{22}) $ denotes, for each $n$, two independent complex coefficients.
Indeed, $\stackrel{\!\!\!\!(3)}{I^{22} }(t)$ being complex, there are no reality conditions relating
$C^{+}_{n} ( I^{22} )$ and $C^{-}_{n} ( I^{22} )$ ( in spite of the fact that 
$\sigma^{+}_{2n}$ and $\sigma^{-}_{2n}$ are related by complex conjugation).

   If we include in Eq.~(\ref{i_3_22_rd}) only the first two complex conjugated fundamental QNM modes,
$\sigma^{+}_{20}$ and $\sigma^{-}_{20}$, we observe that 
$\stackrel{\!\!\!\!(3)}{I^{22}_{\rm ring} }(t)$
contains two arbitrary
complex coefficients $C^{+}_{0} ( I^{22} )$ and $C^{-}_{0} ( I^{22} )$.
These two complex coefficients can be chosen so as to ensure  not only
that $\stackrel{\!\!\!\!(3)}{I^{22}_{\rm plunge } }(t = t_{\rm match} )$
agrees with $\stackrel{\!\!\!\!(3)}{I^{22}_{\rm ring} }(t = t_{\rm match})
= C^{+}_{0} ( I^{22} ) + C^{-}_{0} ( I^{22} )$, but also that the 
(numerically computed\footnote{For a smooth match, one should no longer
use the  quasi-circular approximation, $\dot r \ll r\, \dot \varphi $, when computing
the time derivatives of $\stackrel{\!\!\!\!(3)}{I^{22}_{\rm plunge } }(t )$.})
time derivative $ \frac{d\, \stackrel{\!\!\!\!(3)}{I^{22}_{\rm plunge } }(t ) }{dt}  $
agrees, when $ t= t_{\rm match}$ with 
$ \frac{d\, \stackrel{\!\!\!\!(3)}{I^{22}_{\rm ring }}(t ) }{d \tau} 
= - \sigma^{+}_{20}\, C^{+}_{0} ( I^{22} ) - \sigma^{-}_{20}\, C^{-}_{0} ( I^{22} ) $.
This yields 
\bs
\bea
C^{+}_{0} ( I^{22} ) &=&
\frac{ \biggl [  
\sigma^{-}_{20} \, \stackrel{\!\!\!\!(3)}{I^{22}_{\rm plunge} }(t)
+  \frac{
d\, \stackrel{\!\!\!\!(3)}{I^{22}_{\rm plunge } }(t)}{ dt} 
\biggr ]_{t=t_{\rm match} }
}{ \sigma^{-}_{20} - \sigma^{+}_{20} }\,,
\\
C^{-}_{0} ( I^{22} ) &=&
\frac{ \biggl [ 
\sigma^{+}_{20} \, \stackrel{\!\!\!\!(3)}{I^{22}_{\rm plunge} }(t)
+  \frac{
d\, \stackrel{\!\!\!\!(3)}{I^{22}_{\rm plunge } }(t)}{ dt} 
\biggr ]_{t = t_{\rm match} } }{ \sigma^{+}_{20} - \sigma^{-}_{20} }\,.
\eea
\label{c0_pm_f}
\es
Similarly, we can match, in a once-differentiable ($C^{1}$) manner,
$  
\stackrel{\!\!\!\!(3)}{S^{21}_{\rm plunge}(t)}, 
\stackrel{\!\!\!\!(4)}{I^{31}_{\rm plunge}(t)} $,
and $\stackrel{\!\!\!\!(4)}{I^{33}_{\rm plunge}(t) } $ to ring down moments 
of the form
\bs
\bea
\stackrel{\!\!\!\!(3)}{S^{21}_{\rm ring} }(t) &=&
C^{+}_{0} ( S^{21} )\, e^{- \sigma^{+}_{20}\, \tau }
+ C^{-}_{0} ( S^{21} )\, e^{- \sigma^{-}_{20}\, \tau}
\,,\\
\stackrel{\!\!\!\!(4)}{I^{31}_{\rm ring} }(t) &=&
C^{+}_{0} ( I^{31} )\, e^{- \sigma^{+}_{30}\, \tau }
+ C^{-}_{0} ( I^{31} )\, e^{- \sigma^{-}_{30}\, \tau}
\,,\\
\stackrel{\!\!\!\!(4)}{I^{33}_{\rm ring} }(t) &=&
C^{+}_{0} ( I^{33} )\, e^{- \sigma^{+}_{30}\, \tau }
+ C^{-}_{0} ( I^{33} )\, e^{- \sigma^{-}_{30}\, \tau}\,.
\eea
\label{is_lm_ring}
\es
Each pair of  complex coefficients $ C^{\pm}_{0} ( {\cal M} )$ is then given
as a linear combination of ${\cal M}_{\rm plunge} (t_{\rm match}) $
and $ \frac{ d }{dt} {\cal M}_{\rm plunge} (t_{\rm match}) $
of the type, given by Eq.~(\ref{c0_pm_f}) above.
Finally, we can use the complex conjugation relations, Eq.~(\ref{cc_rules}),
to match, in a $C^{1}$ manner, the remaining required radiative moments
$ \stackrel{\!\!\!\!(3)}{I^{2-2}(t)} $ and $
\stackrel{\!\!\!\!(4)}{I^{3-3}(t)} $ entering Eq.~(\ref{P_flux_f}).
This does not introduce new, independent coefficients as
\be
(-)^m\, C^{\pm}_{0}( I^{lm} ) = C^{\pm}_{0} \left [ ( I^{lm})^* \right ] =
\left [ C^{\mp}_{0} ( I^{lm} ) \right ]^* \,.
\ee
Note also that to match the multipole moments entering the leading order linear momentum flux,
we need to know only two conjugate pairs of complex QNM frequencies, 
namely from Refs.~\citep{CD75} and \citep{KS_LR},
\bs
\bea
\sigma^{\pm}_{20} &= & 0.08896 \pm 0.37367\, i\,,\\
\sigma^{\pm}_{30} &=& 0.09270 \pm 0.59944\, i\,.
\eea
\es
Having so determined continuations of the various relevant multipole moments during the merger phase,
we get an estimate of the final recoil, in the leading-order (quasi-Newtonian)
approximation by integrating, from $-\infty$ to $t_{\rm match} $ and then from 
$t_{\rm match} $ to $+\infty$, the linear momentum balance equation
\be
\frac{ d }{dt} \left (  v^x_{\rm com} +i\, v^y_{\rm com}  \right ) =
-\frac{ 1}{ 336\, \pi } \,
\biggl \{
\sqrt{ 14}\,  \stackrel{\!\!\!\!(3)}{I^{2-2} }\, \stackrel{\!\!\!\!(4)}{I^{31} }
\,
+ \sqrt{ 210}\,  \stackrel{\!\!\!\!(3)}{I^{22} }\, \stackrel{\!\!\!\!(4)}{I^{3-3
} }\,
- 14\,i\,  \stackrel{\!\!\!\!(3)}{I^{2-2} }\, \stackrel{\!\!\!\!(3)}{S^{21} }\,
\biggr \}\,.
\label{LM_FF_L}
\ee
Here the ``radiative moments '' $ \stackrel{\!\!\!\!(l+1)}{I^{lm} } $ and  
$ \stackrel{\!\!\!\!(l+1)}{S^{lm} } $ appearing on the right-hand side (RHS) are given by:
(i) when $ t < t_{\rm match} $ by Eq.~(\ref{i_3_22_S}) and similar ``plunge moments''
$ \stackrel{\!\!\!\!(l+1)}{I^{lm}_{\rm plunge} }, \stackrel{\!\!\!\!(l+1)}{S^{lm}_{\rm plunge} }$
obtained by
differentiating ( while neglecting $ \dot r \ll r\, \dot \varphi$) 
Eqs.~(\ref{is_s_ip}), and 
(ii) when $t > t_{\rm match} $  by  analytical QNM-based ``ring-down moments'' 
$ \stackrel{\!\!\!\!(l+1)}{I^{lm}_{\rm ring} }, 
\stackrel{\!\!\!\!(l+1)}{S^{lm}_{\rm ring} }$,
defined 
by Eqs.~(\ref{is_lm_ring}) above.
The $C^1$ continuity of the moments entering the RHS of Eq.~(\ref{LM_FF_L}) ensures that
the linear momentum flux $\mathcal{F}_\mathbf{P} $ [defined by the RHS of Eq.~(\ref{LM_FF_L})]
is continuous, as well as its  first derivative, across $ t = t_{\rm match}$.

As explained above this $C^1$ matching ensures that one did not introduce leading-order spurious
contributions linked to edge effects.
At the same time,  this matching procedure generically introduces discontinuities
in the second time derivative of  $\mathcal{F}_\mathbf{P} $.
We then see from Eq.~(\ref{i_edge_p_r}) that there will be sub-leading spurious contributions linked to
such discontinuities in 
$\frac{d^2  \mathcal{F}_\mathbf{P} }{dt^2} $.
To study the eventual numerical importance of these higher-order edge effects, we have also
implemented an improved matching procedure 
consisting of including, for each radiative multipole moment,
the first two conjugate pairs of QNMs in Eq.~(\ref{i_3_22_rd}), 
{i.e.} both $n=0$ ( fundamental QNMs) 
and $n=1$ (first excited QNMs). The new required QNM frequencies,
available in Refs.~\citep{KS_LR,CD75}, are
\bs
\bea
\sigma^{\pm}_{21} &=& 0.27391 \pm i\, 0.34671\,,
\\
\sigma^{\pm}_{31} &=& 0.28130 \pm i\, 0.58264
\,.
\eea
\es
As the QNM sums, Eq.~(\ref{i_3_22_rd}), now include $4$ arbitrary complex coefficients 
$ C^{+}_{0},  C^{-}_{0}, C^{+}_{1}, C^{-}_{1} $, we can uniquely determine them by 
demanding that each radiative moments ( say $\stackrel{\!\!\!\!(3)}{I^{22}} $ ), together 
with their first {\em three} numerically computed time derivatives
( $ \frac{d^j \stackrel{\!\!\!\!(3)}{I^{22}} }{ d t^j} \,\, {\rm for}\, j \le 3$)
match across $t_{\rm match}$.

   The matching procedure presented so far was based on considering the leading-order,
quasi-Newtonian, expression, Eq.~(\ref{P_flux_f}), for the flux of linear momentum.
When considering the 2PN correction factor $\tilde F$, as in Eq.~(\ref{Eq.19.old}), we should,
in principle, both include more multipolarities in Eq.~(\ref{P_flux_f}), and PN-corrections 
in the expressions for individual radiative moments, Eqs.~(\ref{is_s_ip}) or Eq.~(\ref{i_3_22_S}).
As we found (see below) that contributions to the recoil due to the ring-down phase are relatively 
small, we decided, for simplicity, to use a less-rigorous, but much simpler, 2PN-level
matching procedure.
The procedure we used consisted in continuing to use the leading-order flux, 
as in Eq.~(\ref{P_flux_f}),
but to ``improve'' the ``brick radiative moments'',
$ \stackrel{\!\!\!\!(3)}{I^{2m}},  \stackrel{\!\!\!\!(3)}{S^{2m}},  \stackrel{\!\!\!\!(4)}{I^{3m}} $,
it contains by multiplying each of them by a factor $\sqrt{\tilde F}$, { e.g.} we 
modify Eq.~(\ref{i_3_22_S}) to
\be
\stackrel{\!\!\!\!(3)}{I^{22}}^{\rm improved}_{\rm  plunge}
= i\,\sqrt{\tilde F}\,
\frac{16}{5}\, \sqrt{10\,\pi}\,\, \eta\,
r(t)^2\,
\omega(t)^3\, e^{-2\,i\,\varphi(t)}\,\,\,\,\,\, ({\rm for  } \,\,\,t < t_{\rm ma
tch} )\,.
\ee
Then  we match each of these ``improved plunge moments'' to a corresponding 
``ring-down''  one, given by a QNM sum of the form, given by Eqs.~(\ref{i_3_22_rd}).
Again this matching can be done in a $C^1$ (2\,\,QNM's) or $C^3$ ( 4 QNM's) manner.

\section{Results}
\label{S5}
Having presented our methodology, let us now discuss the results 
that we obtained, and their interpretation.
Let us consider first the leading-order, quasi-Newtonian approximation,
{ i.e.} Eqs.~(\ref{P_flux_ip}) and (\ref{com_b_ip}), 
together with the leading-order (2 QNM's per moment) matching
to the ring-down phase. We plot in Fig.~\ref{Fig_2_MF_t_A}, for the case of $\eta=0.2$,
the magnitude 
of the linear momentum flux
$| \mathcal{F}^i_\mathbf{P}| $,
together with their two separate components 
$\mathcal{F}^x_\mathbf{P}$ and $ \mathcal{F}^y_\mathbf{P} $,
as functions of time.
The maximum of $|\mathcal{F}^i_\mathbf{P}| $ is reached for $t= t_{\rm max} \simeq 
4149.50  $ 
(which corresponds to $r_{\rm max} \simeq 3.501 $, in an evolution for which
the initial separation when $t=0$ was $r=15$, while the matching to the ring-down phase was done at
$t=t_{\rm match} \simeq 4153.50 $, which corresponds to $r_{\rm match} \simeq 2.933 $.
Note the rather fast (and oscillatory) decay of the individual components of 
$\mathcal{F}^i_\mathbf{P}$ during the ring-down. Indeed, we see 
from Eqs.~(\ref{i_3_22_rd}) to (\ref{is_lm_ring}) that, 
during the ring-down, $ \mathcal{F}^i_\mathbf{P} $ is a sum of contributions
proportional either to  
$e^{-\left ( \sigma_{20}^{ \epsilon_{2}} + \sigma_{30}^{\epsilon_{3}}  \right )\, \tau }  =
e^{-\left ( \alpha_{20}+ \alpha_{30} 
\right )\, \tau } \, e^{-i\,\left (  \epsilon_{2}\, \omega_{20}+ \epsilon_{3}\, \omega_{30} 
\right )\, \tau } 
$ (where $ \epsilon_{2}^2 =1 = \epsilon_{3}^2 $) or to
$e^{-\left ( \sigma_{20}^{\epsilon_{2}} + \sigma_{20}^{\epsilon'_{2}}
\right )\, \tau }  =
e^{- 2\, \alpha_{20}\, \tau } \, e^{-i\, \left (  \epsilon_{2}+ \epsilon'_{2}
\right )\, \omega_{20} \tau }
$ (where $ \epsilon_{2}^2 =1 = {\epsilon'}_{2}^2 $).
From Eqs.~(\ref{i_3_22_rd}) and (\ref{is_lm_ring}), we see that the slowest exponential decay is
$\propto  e^{-2\, \alpha_{20}\, \tau},$ which decays on a characteristic time scale
$\tau_{\rm ring} = \frac{1}{2\, \alpha_{20} } \sim 5.62$.
Though this is significantly smaller than the orbital period near the LSO, note,
however, that this time scale is comparable both to the characteristic time scale
for the variation of $ \mathcal{F}_\mathbf{P}(t) $ near its maximum 
($ \tau_{\rm max} \simeq  7.440 $, see above) and to the inverse of the angular frequency
near the latter maximum ( $ \frac{1}{\omega_{\rm max}} \simeq  \frac{1}{0.1255} \simeq 
7.968
$ ).

   In Fig.~\ref{Fig_3_VC_t_A}, we display the temporal evolution of the recoil velocity, where
we exhibit both its magnitude $| {\mathbf v}^i_{\rm com}(t)|$ and its two separate components
${\mathbf v}^x_{\rm com}(t)$ and ${\mathbf v}^y_{\rm com}(t)$. We see that the maximum
instantaneous recoil velocity is reached {\it after} the maximum of
$| \mathcal{F}_\mathbf{P}|$, and while $| \mathcal{F}_\mathbf{P}| $ has already 
significantly decreased. After having reached its maximum value,
$| {\mathbf v}^i_{\rm com}(t)|$ slightly decreases, in general with some oscillations,
before settling down to its terminal value (see top panel in Fig.~\ref{Fig_3_VC_t_A}). A useful visualization of the evolution
of the recoil  is provided by plotting the ``hodograph'', i.e. a
parametric plot of
the instantaneous recoil velocity vector in the two-dimensional
plane $({\mathbf v}^x_{\rm com}, {\mathbf v}^y_{\rm com})$:
see Fig.~\ref{Fig_traj}.
The behavior of the instantaneous recoil velocity
is easily interpreted in view of the analytical arguments presented above.
Indeed, we have seen in Eq.~(\ref{v_com_estimate}) that, during the plunge, the main contribution 
to the integrated recoil came from the {\it non-adiabatic } character of the 
evolution of $| \mathcal{F}_\mathbf{P}(t)|$ near its maximum.
This led us to estimate that the instantaneous recoil $v_{\rm com}(t)$ was given by the truncated
Gaussian integral, Eq.~(\ref{tr_GI}).
This integral can be expressed by a (complementary) error function
${\rm erfc}(z) = \left ( \frac{2}{\sqrt{\pi}} \right ) \int_{z}^{+\infty} e^{- x^2}\, dx $, with argument
$z =  - \sqrt{ \frac{\alpha}{2}}\, \biggl ( \bar t - \frac{\beta}{\alpha} \biggr )$ where
 $\bar t \equiv t - t_{\rm max} $, and where $ \alpha$ and $\beta $ are defined by
Eq.~(\ref{XY_def}). If, for simplicity, one neglect 
$\epsilon_{\rm max} \ll 1 $, we find that $z \sim -\frac{1}{\sqrt{2}} 
\,\left ( \frac{\bar t}{ \tau_{\rm max}} - i\, \omega_{\rm max} \, \tau_{\rm max} \right )$.
Note that $z$ is shifted (by a quantity of order unity because 
$ \omega_{\rm max} \, \tau_{\rm max} = Q \simeq 0.9337 $ ) in the complex plane.
This shift in the complex plane introduces some modifications to the usual behavior of the
complementary error function in the real domain, which evolves
monotonically from $ {\rm erfc} (+\infty) =0$ when $z=+\infty$, { i.e.}
$\bar t = -\infty $, to
$ {\rm erfc}(-\infty) =2$ when $z=-\infty$, { i.e.}
$\bar t = +\infty $. These modifications are such that the modulus 
$ |{\rm erfc}(z(\bar t))|$ increases from the value $0$ when $ \bar t = -\infty$ to a maximum value of
about $2.05$ when $ \frac{\tau}{\tau_{\rm max}}  \simeq + 1.06$,
before decreasing toward its final value of $2$ when
$ \frac{\tau}{\tau_{\rm max}}  \gg 1 $.
Note also that $|{\rm erfc}(z(\bar t))|$ already reaches the value $\simeq 1.97$ when
$ \frac{\tau}{\tau_{\rm max}} \simeq +1 $. Therefore, most of the integrated effect of the maximum of
$| \mathcal{F}_\mathbf{P}(t)|$ is acquired when $ t - t_{\rm max} \sim + \tau_{\rm max} $.
However, in the case of the evolution depicted in Fig.~\ref{Fig_1_MF_r_t}, one can check that the time
$ t_{\rm max} + \tau_{\rm max} $ corresponds to a radius $r \simeq 2.648 $,
which is (slightly) below $ r \simeq  3$, { i.e.}
after our chosen transition
time to ring-down $ t_{\rm match} \simeq 4153.5 $ (corresponding to
$ r \simeq  2.933$). Therefore, the integrated effect up to
$ t_{\rm match}$ of the non-adiabatic evolution of $| \mathcal{F}_\mathbf{P}(t)|$ near its maximum will be slightly smaller than
the total integrated effect considered above.

In addition, the transition from  $  \mathcal{F}^{\rm plunge}_\mathbf{P}(t)$
to $  \mathcal{F}^{\rm ring}_\mathbf{P}(t)$
across $ t_{\rm match}$ introduces a new source of non-adiabaticity
\footnote{The non-adiabatic character of the transition between plunge and ring-down
shows up particularly in the fact that one passes from a quasi-monochromatic (chirping)
form $  \left [ \mathcal{F}^{x}_\mathbf{P}(t)
+  i\, \mathcal{F}^{y}_\mathbf{P}(t) \right ]_{\rm plunge} = a(t)\, e^{i\, \varphi(t)} $
to a non-monochromatic form containing both (decaying) positive and negative frequencies:
$ e^{\alpha\, \tau} \, e^{\pm i\,\omega\, \tau} $.}.
Fig.~\ref{Fig_3_VC_t_A} shows that the ring-down behavior can introduce some oscillations in
$|{\mathbf v}_{\rm com} |$, and tends to {\it decrease} the value of 
$|{\mathbf v}_{\rm com} |$ reached after passing the maximum of
$ | \mathcal{F}_\mathbf{P} | $. However, one sees on the plot that the effect of ring-down is relatively
small compared to the main contribution to $|{\mathbf v}_{\rm com} |$ acquired by passing 
over the maximum of $ |\mathcal{F}_\mathbf{P}(t)| $.
The relative smallness of the ring-down contribution to $|{\mathbf v}_{\rm com} |$
can also be checked analytically. Indeed, this ring-down contribution is
given by a sum of integrals of the form
$ \int_{0}^{+\infty} d\tau\, {\cal C}\,
e^{- \left( \alpha + i\, \omega \right ) \,\tau}
= \frac{{\cal C}}{  \alpha + i\, \omega  }$.
One can then relate this sum of integrals to the value of 
$ | \mathcal{F}_\mathbf{P} | $ at the moment of the matching.
One then checks that, because the transition occurs while $ | \mathcal{F}_\mathbf{P} | $
has already significantly decreased from its  maximum value, the ring-down integral will be
significantly smaller than the value acquired by passing over the maximum of
$|\mathcal{F}_\mathbf{P}(t) |$. 

  In Figs.~\ref{Fig_4_MF_t_B} and \ref{Fig_5_VC_t_A},
we study the effect of demanding a smoother transition between plunge and ring-down, namely
a $C^3$ one ( with two conjugate pairs of QNM's per multipole moments) instead of the 
$C^1$ one ( one pair of QNM's) used in the 
Figs.~\ref{Fig_2_MF_t_A} and \ref{Fig_3_VC_t_A}.
As we see, though the effect is not 
negligible (and can introduce some extra oscillations in ${\mathbf v}^i_{\rm com}$),
it has a relatively minor effect on the final recoil velocity.
More precisely, we find that (for $\eta=0.2$) 
$v^{\rm terminal}_{\rm com}|_{\rm 2\,\, QNM}  
\simeq 51.05\,\, {\rm km}/{\rm s} $, 
while 
$v^{\rm terminal}_{\rm com}|_{\rm 4\,\, QNM}  
\simeq  54.19 \,\, {\rm km}/{\rm s} $.

    Let us now consider the impact on 
$v^{\rm terminal}_{\rm com}$  of the higher-order PN corrections to $ \mathcal{F}_\mathbf{P}$,
{i.e.} the effect of the factor $\tilde F$ in Eq.~(\ref{Eq.19.old}).
The definition of $\tilde F$, given by Eq.~(\ref{Ftilde}),
depends on the definition of $F(v)$\footnote{For simplicity, 
in view of the closeness of $\psi(r, p_{\varphi})$ to $1$, we shall not
contemplate other definitions of $\tilde F$ based, {\it e.g.}, on expanding {\it both }
$\psi$ and $F$ in PN series before, eventually re-summing the PN expansion of $\tilde F$.}.
We have discussed above various ways of estimating $F$:
one can use straightforward ``Taylor approximants'', ( {\it e.g.} $ F^{\rm T}_{1PN} \equiv
1 + F_2\, v^2 $), or, instead, some ``Pad\'e'' ones
( {\it e.g.} $ F^{\rm P}_{1PN} = \frac{1}{ 1 - \frac{v}{v_{\rm pole} }} \, 
\frac{1}{  1 + \frac{c_1\,v} {  1 + c_2\,v  }   } $ ) .
Actually, the 1PN-accurate Taylor approximant $ F^{\rm T}_{1PN} \equiv
1 + F_2\, v^2 $ is not an
acceptable approximation for the study of the recoil. Indeed, when $\eta =0.2$, one finds that
$ F^{\rm T}_{1PN} $ becomes {\it negative} for $v \ge 0.421 $.
As $v_{\varphi} = \omega\, r$ is about $0.44$ at the maximum of 
$ \mathcal{F}_\mathbf{P}$, the most important domain of values for $v_{\varphi}$ to estimate
the recoil would correspond to such a physically incorrect negative value for $ F^{\rm T}_{1PN} $.
Though the corresponding Pad\'e approximant $ F^{\rm P}_{1PN}$ stays positive,
it is also unphysical in that it takes values of order $10$ in the relevant range of values 
for $v$. Therefore, we shall only consider the higher-order PN approximants:
$ F^{\rm T}_{1.5PN}$, $ F^{\rm T}_{2PN}$, $ F^{\rm P}_{1.5PN}$, and $ F^{\rm P}_{2PN}$.
In Table~\ref{Tab_1}, we present (for $\eta=0.2$), the values of $|{\mathbf v}_{\rm com}|$,
${\mathbf v}^{x}_{\rm com}$, and ${\mathbf v}^{y}_{\rm com}$ for various 
approximants and at various stages of the inspiral or merger.
We observe that changing the approximant for $\tilde F$ has a substantial effect on
$v_{\rm com}$. 
In particular, the final recoil varies between $v^{\rm terminal}_{\rm com} \simeq 50\,\,
{\rm km}/{\rm s}$
when using a 2PN accurate Pad\'e approximant, and $v^{\rm terminal}_{\rm com} \simeq 74 
\,\,{\rm km}/{\rm s}$
with a 2PN accurate Taylor approximant (the quasi-Newtonian estimate being 
$v^{\rm terminal}_{\rm com} \simeq 51 \,\, {\rm km}/{\rm s} $). 
In agreement with the approximate analytical estimate derived above,
one can check that $v^{\rm terminal}_{\rm com}$ varies in direct proportion
to the value taken by the PN factor $\tilde F$ at $t= t_{\rm max}$,
{ i.e.} when $|\mathcal{F}_\mathbf{P}(t) |$ reaches its maximum value.
Indeed, neglecting the effect of $\psi$ in Eq.~(\ref{Ftilde}) 
[$\psi (t_{\rm max}) \sim 1.007 $], the variation
of $v^{\rm terminal}_{\rm com}$ with $F$ is well describable by the variation of
$F(v_{\rm max})$ with $v_{\rm max} \simeq 0.439 $.
Indeed, we have the following  values for $F$, namely,
$F_N (v_{\rm max}) =1 $, $ F^{\rm T}_{1.5PN} (v_{\rm max})
\simeq 1.33 $, $ F^{\rm T}_{2PN}(v_{\rm max}) \simeq 1.35,
F^{\rm P}_{1.5PN}(v_{\rm max}) \simeq 1.27 $, 
and $ F^{\rm P}_{2PN}(v_{\rm max}) \simeq 0.93 $ which are well correlated with the results
listed in Table~\ref{Tab_1}.
The significant difference between $ F^{\rm T}_{2PN}(v_{\rm max})$
and $F^{\rm P}_{2PN}(v_{\rm max})$ illustrates again the poor convergence of the 
successive PN contributions.
As argued earlier, one would expect [by analogy with convergence of the energy-flux function
$F_E(v)$] that the $2.5$ PN-accurate Pad\'e approximant would yield a better answer.
In absence of information concerning the 2.5 PN level, we shall use $F^{\rm T}_{2PN}$
as our best answer [in view of the behavior of $F^{\rm T}_{E_{2PN}}$ \cite{DIS98}],
but keep in mind the probability of a significant error bar around it.

  In Table~\ref{Tab_2}, we study the influence of another parameter in
our methodology:
the precise choice of the transition radius between plunge and ring-down.
We consider the standard Schwarzschild light ring, namely
$r=3$, as our default value. In Table~\ref{Tab_2}, we explore the effect on $v^{\rm terminal}_{\rm com} $
of changing the matching radius $r_{\rm match}$ by $\pm 20 \%$ of its default value. 
As this Table shows, the value of
$v^{\rm terminal}_{\rm com} $ is mildly sensitive to the precise choice
of transition radius.

 Up to now, we have only focused on a specific symmetric mass ratio ($\eta=0.2$) which,
in view of the analytical estimate, given by Eqs.~(\ref{v_com_estimate})
and (\ref{ana_est_v_com}), is expected to yield the maximum possible recoil.
In Table~\ref{Tab_3}, we consider several
different values of $\eta$, namely $\eta=0.05, 0.1, 0.2,$ and 
$0.24$ and compute the corresponding {\it scaled} 
terminal recoil ${\hat v}_{\rm com} = \frac{v_{\rm com}}{f(\eta)}$,
where $f(\eta)= \eta^2\, \sqrt{1 - 4\,\eta} $.
As expected, after scaling out the function $f(\eta)$, the recoil depends only weakly on $\eta$.
We can analytically approximate the $\eta$-dependence of $\hat v_{\rm com}$ by a second-order polynomial
$P(\eta) = a + b\, \eta + c\, \eta^2$.
Normalizing $a,b$ and $c$ so that $P(0.2) =1$, we find
a reasonable fit\footnote{Actually, to get a good fit
to the data in Table~\ref{Tab_3} one needs a third-order polynomial,
namely: $P_3(\eta) =  1.112 - 1.78\,\eta + 10.3\, \eta^2 \,
- 21\,\eta^3.$}
for
\be
P(\eta) =  1.0912 - 1.04\,\eta + 2.92\, \eta^2 \,.
\ee

   Finally, putting together the various informations we have obtained
  above, we can summarize our ``best bet'' estimate for the final recoil
associated with the coalescence of binary black holes of symmetric
mass ratio $\eta$ as

\be \label{finalestimate}
v^{\rm final}_{\rm com} \simeq
73.5\, \frac{ \tilde F_{\rm max} }{1.35}\, \hat f(\eta)\, {\rm km/s}\,,
\ee
where $\hat f(\eta) = \frac{ \eta^2\, \sqrt{1 - 4\,\eta} }{ 0.0178885}\,
\left ( 1.0912 - 1.04\,\eta + 2.92\, \eta^2 \right ) $.
The fiducial
value $ 1.35 $ used above for scaling $\tilde F_{\rm max}$ is the
prediction made by the 2PN Taylor approximant around $r_{\rm max}
\simeq 3.50$, i.e. at the moment where the modulus of the
linear momentum flux is maximum. Note that the proportionality
of the final result to $\tilde F_{\rm max}$ is only  approximate
because the presence of the correction factor $\tilde F(t)$
changes not only the height of the maximum of
 $|\mathcal{F}_\mathbf{P}|$ but also affects the shape of
 $|\mathcal{F}_\mathbf{P} |(t)$ and thereby the quality factor $Q$ etc.
The above estimate is plotted as a function of $q=\frac{m_2}{m_1}$ in Fig.~\ref{Fig_7}.

 \section{Discussion}

 The main fruit of the present study is the fact that we have
 delineated, often by means of analytical arguments, the relative
 importance of several different physical effects in determining
 the magnitude of the final recoil velocity $v^{\rm final}_{\rm com}$.
We have emphasized that the value of $v^{\rm final}_{\rm com}$
is essentially determined by a brief period during the
orbital evolution when the integrand
 of the oscillatory integral (\ref{v_f_int})
yielding $v^{\rm final}_{\rm com}$ varies in a non adiabatic
manner: ${\dot a}/a \sim \dot \varphi$. We have found that
this non-adiabatic evolution is confined to a small neighborhood
of the moment, during the plunge, where the modulus of the
linear momentum flux $|\mathcal{F}_\mathbf{P} |(t)$
[i.e. the amplitude $a(t)$ in the integral
(\ref{v_f_int})] reaches a maximum.
The good news is that it seems that this maximum takes place during the quasi-circular ``plunge phase'', i.e. during a phase where
the radial kinetic energy is significantly smaller
than the azimuthal kinetic energy. Indeed, the ratio
$ {\cal R} \equiv g^{RR} p_R^2 / g^{\varphi \varphi} p_{\varphi}^2$
between ``radial'' and ``azimuthal'' kinetic energies is found to
take the value ${\cal R}_{\rm max} \simeq 0.135  $ at $ r_{\rm max}$.
[Note again, that even at the light ring, $r \simeq 3$ this ratio remains small,
namely, ${\cal R}_{\rm lr} \simeq 0.281  $].  

 This ``burst'' of linear momentum flux also occurs
 slightly before the merger phase which we view as
taking place when the (adimensionalized) radial distance 
$r$ gets smaller than about $3 $. As was argued in Ref.~\cite{BD00} the quasi-circular
``plunge'' phase ($ 3< r <6$)
is a priori amenable to analytical description
within the EOB approach.
And we have indeed checked
that various different ways of completing the EOB approach
by a suitable matching to a subsequent ($r < 3$) ring-down
description of the merging of the two black holes did not
affect much the recoil velocity acquired after passing over
the maximum of $|\mathcal{F}_\mathbf{P} |(t)$. We have also
verified that various other physical ingredients of the model
(such as: the representation of the damping force during the
plunge, the choice of matching point, the number of quasi-normal modes, 
$\cdots$) had a rather mild effect on the final recoil.

However, the bad news is that when $|\mathcal{F}_\mathbf{P} |(t)$
reaches its maximum value, the azimuthal kinetic energy
contribution $p_{\varphi}^2/r^2$ in the Hamiltonian equations,
Eqs.~(\ref{eob_eq_s}), as illustrated in Fig.~\ref{RKE_AKE_fig}, is
of order unity ( $p_{\varphi}^2/r^2_{\rm max} \sim 1$), i.e. comparable to the constant term ($=1$), which plays the role of the ``squared rest mass'' term in the Hamiltonian ($ H \sim \sqrt{{\bf p}^2 + m^2}$).
 This situation contrasts with the one near
 the LSO, where one has
$p_{\varphi}^2/r^2_{LSO} \sim 0.25 $ which is significantly smaller
than unity. In other words, the orbital motion near the LSO is still
``non relativistic'' (by a thin margin), while the recoil is generated
 when the
orbital motion becomes mildly relativistic (in the sense ${\bf p}^2 \sim m^2$). What further complicates
the matter is that the orbital motion does not follow the usually
considered sequence of circular orbits, so that we cannot use
the standardly assumed relativistic version of
Kepler's third law relating the
angular velocity to the radius. In the body of the paper, we
have used the ``quasi-Newtonian'' expression for the linear
momentum flux (in EOB coordinates)
 as a guideline to select a ``best bet'' modeling of $\mathcal{F}_\mathbf{P} (t)$ during the plunge.
 We have already seen in Table~\ref{Tab_1}
that a very significant source
of uncertainty in the magnitude of $\mathcal{F}_\mathbf{P} (t)$
concerns the inclusion of post-Newtonian corrections in it.
Depending on whether one uses the straightforward ``Taylor-expanded''
2PN correction \cite{BQW05}, or one of its Pad\'e-resummed versions,
one gets  a multiplicative
factor varying between $0.92$ and $1.35$ in $v^{\rm final}_{\rm com}$.
 However, this uncertainty is only
a {\it lower limit} to the total uncertainty currently attached to
the description of the relativistic effects during the plunge.
We can see hints of a larger uncertainty by comparing our
EOB-based treatment (which did not assume the validity
of the standard Kepler law during the plunge) to a treatment
similar to the one advocated in Ref.~\cite{BQW05} (which did
implicitly assume the continued validity of Kepler's law).
 To explore this issue, and also
to understand the relation between our estimate and the significantly
larger one obtained in Ref.~{\cite{BQW05}, we have  estimated the recoil
following from using the functional form
(\ref{BQWflux}) for $\mathcal{F}_\mathbf{P} $, instead of our
 2PN-corrected, quasi-Newtonian expression
(\ref{Eq.19.old}). As we discussed above, these two different
functional forms can be matched above the LSO
[modulo the suitable definition
of the 2PN-correcting factor $\tilde F$ in Eq.~(\ref{Eq.19.old}),
see Eq.~(\ref{Ftilde})], by using the relativistic
Kepler law (\ref{kepler1}). On the other hand, Eq. (\ref{kepler1})
gets strongly violated below the LSO, and as a consequence
the two different functional forms,
Eqs.~(\ref{BQWflux}) and (\ref{Eq.19.old}), 
a priori lead to quite different time evolutions for
$\mathcal{F}_\mathbf{P}(t) $. In keeping with our general philosophy,
it is useful to understand analytically the difference between the
two basic corresponding ``quasi-Newtonian'' prescriptions [obtained
by neglecting the 2PN factor $F$ in Eq.~(\ref{BQWflux}) and the
corresponding 2PN factor $\tilde F$ in Eq.~(\ref{Eq.19.old})].
In other words, let us discuss the effect of replacing our
basic ``fiducial'' quasi-Newtonian momentum flux
${\cal F}_{r,\omega} = r^5 \omega^7$ by
${\cal F}_{v_{\omega}} =  v_{\omega}^{11} = (\omega)^{11/3}.$
Let us call $K \equiv \omega^2 r^3$ the quantity which is (approximately)
constant when Kepler's law is satisfied. It is then easy to see
that the ratio between the two prescriptions reads:
${\cal F}_{v_{\omega}}/{\cal F}_{r,\omega} = K^{-5/3}$.
We have shown above how to write an approximate evolution equation
for the angular frequency during the plunge, namely:
$\omega \propto A(r)/r^2$. This entails that $K$ varies
during the plunge approximately as $K \propto A^2(r)/r$.
It can be analytically checked that $K$, or better ${\hat K} = \psi K$
with $\psi$ defined by Eq.~(\ref{psieqn}) (augmented
by the needed $p_r^2$ terms), is equal to 1, and
has a horizontal derivative, at the LSO, and \emph {decreases}
monotonically  when $r < r_{LSO}$, to reach zero when
$r$ tends to the ``horizon'' [$A(r_{horizon}) = 0$].
This behavior is illustrated in Fig.~\ref{Fig_2}.
As a consequence $K$ gets (significantly) smaller than 1 below
the LSO, so that ${\cal F}_{v_{\omega}}/{\cal F}_{r,\omega} = K^{-5/3}$
is significantly larger than 1 during the plunge. More precisely,
we can, as above, study the evolution of ${\cal F}_{v_{\omega}}(t)$
by writing that it approximately varies like
${\cal F}_{v_{\omega}} = [\omega(r)]^{11/3} \propto
[A(r)/r^2]^{11/3}$. It can be seen that this function of $r$
has a maximum at $r'_{\rm max} \approx 2.879$, and that the
value of the function at its maximum is more than
{\em twice} higher than the maximum  that ${\cal F}_{r,\omega}
= r^5 \omega(r)^7$ had at  $r_{\rm max} \approx 3.443$.
This increase in the maximum value of the linear momentum flux
is further compounded by significant changes in the shape
of the maximum (notably the value of
the crucial quality factor $ Q \equiv \omega_{\rm max} \, \tau_{\rm max} $ introduced above).
 We therefore see that using
a momentum flux given by Eq.~(\ref{BQWflux}) instead of
Eq.~(\ref{Eq.19.old}) will more than {\em double} the
value of $v^{\rm final}_{\rm com}$[see precise numbers below].
Note  that the large change in the predicted value for
$v^{\rm final}_{\rm com}$ that we just discussed
concerns only the ``leading-order quasi-Newtonian''
expression that one chooses to employ during the plunge.
It has to be further compounded by the uncertainty linked
to the resummation of the 2PN correction
factor $F$ or $\tilde F$ (which brings
a multiplicative uncertainty factor of order $1.5$).

Actually, the fact that the location of the maximum of the
momentum flux ($r'_{\rm max} \approx 2.879$)
is slightly {\em below} the light ring in the case of Eq.~(\ref{BQWflux})
brings a further complication. Indeed, in that case we cannot trust
our simple ``Gaussian integral estimate'' Eq.~(\ref{v_com_estimate})
which assumed that one was integrating over the maximum of
$|\mathcal{F}_\mathbf{P}| (t)$. The matching to the
subsequent ring-down behavior could a priori have a significant impact.
To resolve this issue we have run numerical simulations
similar to what we use in the body of the paper (with
EOB dynamics as defined above, and two Quasi-Normal-Modes
matching at $r=3$) except that the right-hand side of
Eq.~(\ref{Eq.19.old}) was replaced by Eq.~(\ref{BQWflux}).
Our results are displayed in Fig.~\ref{Fig_8} and 
listed 
in the last two rows of Table~\ref{Tab_1}.
In agreement with the simple analytic arguments above, we do find
final recoil velocities that are more than twice larger
when using Eq.~(\ref{Eq.19.old}) (with a corresponding
factor $\tilde F$). 
For instance, from Fig.~\ref{Fig_8}, we find that the terminal recoil is $\sim 172\,{\rm km/s}$
which is more than double the value that we read in the Table~\ref{Tab_1} for a
2PN accurate $\tilde F$.
Furthermore, using Eq.~(\ref{BQWflux})
as it stands we observe, for the optimal case $\eta =0.2$ and while
terminating the EOB evolution at $r\simeq 2$ without doing QNMs matching,
$v^{\rm final} _{\rm com} \simeq 243 \,{\rm km/s} $. On the other hand,
if we do not include the 2PN correction factor $F$ in
Eq.~(\ref{BQWflux}) we obtain, again for $\eta =0.2$, a final
recoil $ \simeq 135\, {\rm km/s}$ 
at $r \simeq 2$.
These results are roughly consistent
with the results of Ref.~\cite{BQW05}, and confirm our diagnostics
that the main physical origin of the integrated recoil
is the rather well-localized
``burst'' in linear momentum flux occurring during the plunge.

One might view the large difference between
 the two models of momentum flux, Eq.~(\ref{Eq.19.old}) versus Eq.~(\ref{BQWflux}),
in several different ways. One way would be to
say that, as we have seen,
it is not justified to continue assuming (as is implicitly
done in Eq.~(\ref{BQWflux})) the validity of Kepler's law during the plunge,
and therefore that the corresponding prediction for $v_{com}$ is
definitely too large\footnote{Because, as we have just seen,
the known violation of Kepler's third law, i.e. $K < 1$,
is the root of the difference between the two estimates.}.
On the other hand, we have emphasized above that near the crucial
maximum of radiation of linear momentum,  the
orbital motion becomes mildly relativistic ($  {\bf p}^2 \sim m^2$, {i.e.} $(v/c)^2 \sim 0.5$)
and, in addition, more complicated than the
quasi-circular and quasi-Keplerian cases studied
so far in detail in analytical gravitational wave research.
Therefore, one might also say that the large difference between the
two proposed extrapolations for the momentum flux just reflects
our ignorance of what is the correct momentum flux in such
a relativistic situation. We do tend to
 think that our ``best bet'' estimate, Eq.~(\ref{finalestimate}),
is probably closer to the truth, but
we cannot provide any proof of this belief,
nor can we presently define an ``error bar'' around our
preferred estimate (\ref{finalestimate}). One way to estimate
an ``error bar'' around (\ref{finalestimate}) would be to study the
effect of using a 3PN-accurate EOB dynamics, and/or to include all the
contributions proportional to $\dot{R}$ and $\ddot{R}$ (which were
neglected here) in the momentum flux. On the other hand, we consider
it likely that
the results quoted above (and listed in the last two rows of
Table~\ref{Tab_1}) coming from the Kepler-law based
  Eq.~(\ref{BQWflux}) furnish
 an {\em upper bound} on the correct recoil.

 We conclude that
none of the current analytical-based estimates of the total
recoil are reliable.
From Table~\ref{Tab_1}, we see that, depending on the analytical
representation used for the linear momentum flux during the plunge, we get a final
recoil velocity which varies in the range $49-172$ km/s.
This ``uncertainty range'' in theoretical predictions is illustrated
in Fig.\ref{Fig_7}.
We view this current theoretical uncertainty
 as a strong motivation for devoting future work to the
specific issue of selecting a reliable\footnote{By ``reliable''
we mean here ``accurate within 50 \%'' or so. The
current uncertainties in analytical estimates span a factor $\sim 5$.}
 analytical model of linear
momentum flux during the plunge. There are several avenues one
could use towards this aim. One might generalize the work
of Ref.~\cite{BQW05} by keeping the terms proportional to
${\dot r}^2$ or $p_r^2$, and by making a minimal use of
 Kepler's third law\footnote{For instance, one might use
 Kepler's law only at the level of the derivation of the multipole
 moments, and not use it anymore when time-differentiating them
 along plunging orbits. One might also use resummation techniques,
 notably by using the resummed quasi-Schwarzschild coordinates
 that enter the EOB formalism.}. Another avenue is
 to compare analytical and numerical results in the case of a
 test-particle, $\eta \ll 1$, plunging into a Schwarzschild
 black hole. One might also try to compare analytic
 and numerical results in the case of full 3-d simulations
 of coalescing black holes.\footnote{After the submittal
 of this paper, remarkable progress
in numerical relativity has allowed Baker et al. \cite{kick_NR} to report
on the first accurate numerical calculation of the recoil
velocity of two non-spinning black holes with $\eta =0.24$.
Their result ($v_{\rm recoil}=105\,$km/s) corresponds,
if we use the scaling function $\hat f (\eta)$, defined after 
Eq.~(\ref{finalestimate}),
to a maximum recoil of $161.5$km/s. This value falls within
the range of analytical estimates summarized in Table \ref{Tab_1}.
More precisely, the last row of Table \ref{Tab_1} shows that the
momentum flux Eq.~(\ref{BQWflux}) predicts a recoil of 110 km/sec
when $\eta =0.24$.
We note also that the time evolution, during the coalescence,
of the magnitude of the instantaneous recoil velocity
showed in Fig.~1 of \cite{kick_NR} is in good qualitative agreement
with our analytical predictions (see Figs.~\ref{Fig_3_VC_t_A}, \ref{Fig_5_VC_t_A}
or \ref{Fig_8}).}
However, one should keep in mind
that the linear momentum flux is a sub-dominant effect
 in the gravitational wave emission which can easily get drowned
 in the ``noise'' associated, for instance, with the presence
 of residual incoming radiation in the initial data.
 Let us note in this respect, that the situation is a priori
 much better in the (more urgent) problem of the modeling
 of gravitational radiation (linked to the dominant energy flux)
 from coalescing black holes. In this case most of the signal-to-noise
 ratio is linked to the train of waves emitted in the last few orbits
 before crossing the LSO. The EOB formalism was conceived
  for dealing with this problem, and the current study should
 not be interpreted as casting
 any serious doubt on previous studies \cite{BD00, BCD05} which relied
 mainly on rather robust analytic features of the EOB approach.

   If our `best bet' estimate, Eq.~(\ref{finalestimate}), is confirmed, it might
have significant astrophysical consequences. 
For instance, a low value for the terminal recoil will have implications for the formation of 
massive black holes at high redshifts \cite{YJ04}, which may influence various 
event rates for binaries involving supermassive black holes that LISA may observe.
Finally, a recent surge in astrophysical investigations that
probe  consequences of
the recoil indicates that it is desirable to know the dependencies of recoil on
the orbital configurations of coalescing black hole binaries \cite{pap_recoil}.
Therefore, it will be also important, in the near future, to extend our
approach to the study of the recoil associated with the coalescence of 
spinning black holes in inspiralling
eccentric orbits.

\begin{acknowledgments}
We are grateful to Steven Detweiler, Sai Iyer and Kostas Kokkotas
for informative communications.
AG thanks Gerhard Sch\"afer for discussions and encouragements.
TD thanks the European Research and
Training Network `Forces Universe' (contract number MRTN-CT-2004-005104)
for partial support.
AG gratefully acknowledges the financial support of
the Deutsche Forschungsgemeinschaft (DFG) through SFB/TR7
``Gravitationswellenastronomie''. He also thanks IHES for hospitality during early stages of the work.

\end{acknowledgments}

\newpage

\begin{table}[t]
\begin{center}
\caption{
Values of $|{\mathbf v}_{\rm com}|$, ${\mathbf v}^{\rm x}_{\rm com}$ and
${\mathbf v}^{\rm y}_{\rm com}$, recoil velocity and its $x$ and $y$ components in ${\rm km}\,
{\rm s}^{-1}$ for $\eta =0.2$ at various stages of coalescence for
the momentum flux Eq.~(\ref{Eq.19.old}) with different $\tilde F$'s.
The last two separated rows (corresponding to $\eta=0.2$ and $\eta=0.24$) show the recoil estimates obtained by
using, instead of  Eq.~(\ref{Eq.19.old}), the (Kepler-law-based)
momentum flux Eq.~(\ref{BQWflux}).
}
\label{Tab_1}
\begin{tabular}{l||r|r|r}
$$& $r_{\rm LSO} \simeq 6.00 $ & $r_{\rm match} \simeq 3.00$  & $t \rightarrow + \infty$ \\
\hline\hline
1.5 PN Taylor $\tilde F$ &22.09, 18.05,12.73 
& 82.84,51.73,-82.68 
& 72.41,12.18,-71.37
\\ \hline
2 PN Taylor $\tilde F$ &22.34,18.25,1.29
&  84.07, 5.23,-83.91
& 73.52,12.31,-72.48
\\ \hline
$\tilde F =1 $ & 18.50,15.41,10.24
& 59.83,4.29,-59.67
&51.05,10.29,-50.00
\\ \hline
1.5 PN Pad\'e $\tilde F$& 20.69,16.91,11.92 
& 79.97,5.11,-79.81
&  70.19, 11.68,-69.21
\\ \hline
2 PN Pad\'e $\tilde F$ & 16.20,13.42,9.08
&  56.68, 3.96,-56.54
&  49.02,9.15,-48.16
\\
\hline\hline
Eq.~(\ref{BQWflux}), $\eta =0.2$ & 23.39, 19.34,13.16
& 179.88, 78.54, -161.82
& 171.55, 136.64,-103.73
\\ 
\hline
Eq.~(\ref{BQWflux}), $\eta=0.24$ & 15.47, -5.03,-14.62
& 118.67,-85.74,82.05
& 109.64,-103.45,36.34
\\ 
\end{tabular}
\end{center}
\end{table}

\begin{table}[c]
\begin{center}
\caption{ 
Effect of changing the transition point where merger phase goes to QNM ringing 
on terminal $|{\mathbf v}_{\rm com}|$ for $\eta =0.2$ and $\tilde F = 1$, when
changing the canonical value, $r_{\rm match} \simeq 3$, by $\pm 20 \%$.
}
\label{Tab_2}
\begin{tabular}{l||r|r|r}
& $|{\mathbf v}_{\rm com}|$
&${\mathbf v}^{\rm x}_{\rm com}$ 
&${\mathbf v}^{\rm y}_{\rm com}$
\\ \hline\hline
$r_{\rm match} \simeq  3.6 $& 44.98 & -8.05 & -44.26
\\ \hline
$r_{\rm match} \simeq 3.00 $ & 51.05 & 10.29 & -50.00
\\ \hline
$r_{\rm match} \simeq 2.4 $& 58.21 & 12.60 & -56.83
\\ 
\end{tabular}
\end{center}
\end{table}

\begin{table}[b]
\begin{center}
\caption{
Dependence of scaled terminal recoil velocity  $
\frac{ 
|{\mathbf v}_{\rm com}|}{ \eta^2\, \sqrt{1- 4\,\eta}}$
on $\eta$ for $r_{\rm match} \simeq 3.00$ and $\tilde F = 1$.
}
\label{Tab_3}
\begin{tabular}{l||r}
& In km\,s$^{-1}$
\\ \hline\hline
$ \eta =0.24  $& 2817.63
\\ \hline
$ \eta =0.2  $& 2853.81
\\ \hline
$ \eta =0.1  $& 2901.29
\\ \hline
$ \eta =0.05  $& 2987.44
\\
\end{tabular}
\end{center}
\end{table}

\clearpage

\begin{figure}[c]
\begin{center}
\resizebox{9cm}{!}{\includegraphics{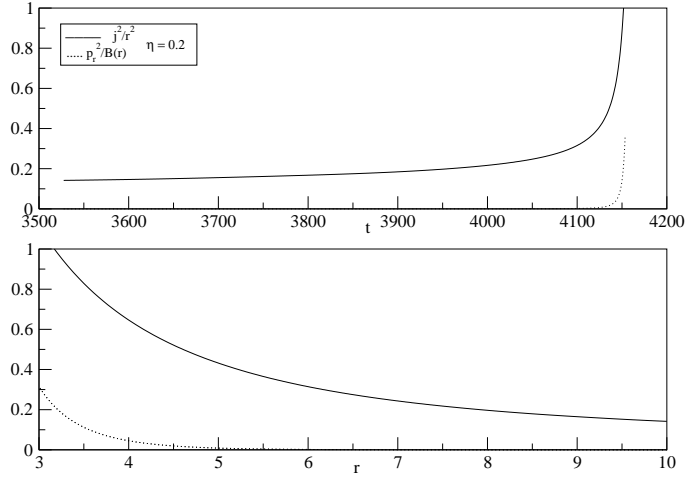}}
\caption{
Plots of `azimuthal' and `radial' kinetic energies as functions of 
coordinate time $t$ and radial separation $r$ during the 
late inspiral and plunge for a $\eta=0.2$ binary. The initial orbital separation,
when $t=0$, was $r=15$, and the curves were terminated at
$t \simeq 4153$, corresponding to $r=3$.
The plots show that $j(t)^2/r^2$ dominates $p_r(t)^2/B(r)$ during the
entire EOB evolution, including the plunge (down to $r=3$).
\label{RKE_AKE_fig}}
\end{center}
\end{figure}


\begin{figure}[c]
\begin{center}
\resizebox{10cm}{!}{\includegraphics{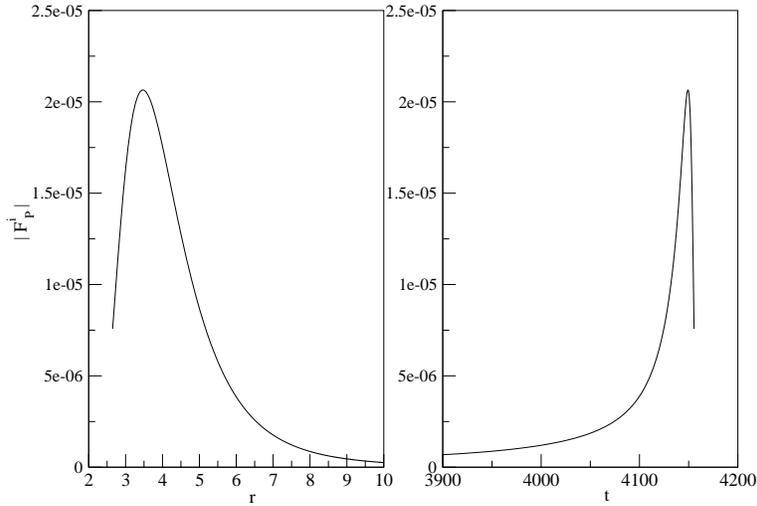}}
\caption{Magnitude of the linear momentum flux (in the quasi-Newtonian
approximation) as a function of $r$ (left panel)  and $t$ (right
panel) during
the late inspiral and plunge for a $\eta=0.2$ binary, whose initial orbital separation,
when $t=0$, was $r=15$. We terminated the plunge arbitrarily around $r \simeq 2.65$.
\label{Fig_1_MF_r_t}}
\end{center}
\end{figure}

\clearpage

\begin{figure}[c]
\begin{center}
\resizebox{9cm}{!}{\includegraphics{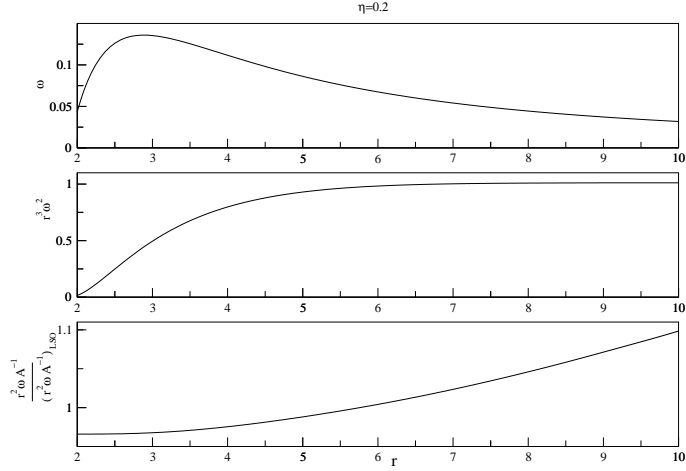}}
\caption{
Plots of $\omega, r^3\,\omega^2 $ and
$ \frac{ r^2\, \omega\, {\rm A}^{-1} }{ \left ( r^2\, \omega\, {\rm A}^{-1} \right )_{\rm LSO} } $
in terms of $r$ resulting from the EOB evolution, given by Eqs.~(\ref{eob_eq_s}), for $\eta =0.2$ binary.
The panels clearly demonstrate that the orbital frequency evolves
differently during the late inspiral and the subsequent plunge. Note
in particular the strong decrease of the ``Kepler combination''
$K \equiv r^3\,\omega^2 $
during the plunge.
\label{Fig_2}}
\end{center}
\end{figure}


\begin{figure}[c]
\begin{center}
\resizebox{10cm}{!}{\includegraphics{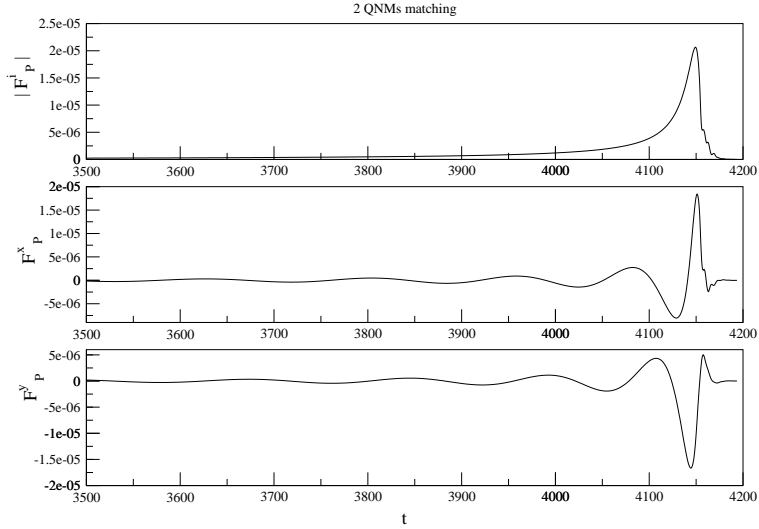}}
\caption{ Plots for the magnitude, $x$ and $y$ components of the linear momentum flux (quasi-Newtonian approximation) versus $t$
during the coalescence of a binary with $\eta=0.2$. For this figure, the ring-down phase is 
described by 2 QNMs and orbital separation of the binary, when the numerical evolution began,
was $r=15$.
\label{Fig_2_MF_t_A}}
\end{center}
\end{figure}

\clearpage

\begin{figure}[c]
\begin{center}
\resizebox{10cm}{!}{\includegraphics{A_FigVQA.eps}}
\caption{ Temporal plots for the magnitude, $x$ and $y$ components of
recoil velocity for the binary configuration described in 
Fig.~\ref{Fig_2_MF_t_A} ($\eta =0.2$, quasi-Newtonian flux, 2 QNM's).
\label{Fig_3_VC_t_A}}
\end{center}
\end{figure}

\begin{figure}[c]
\begin{center}
\resizebox{10cm}{!}{\includegraphics{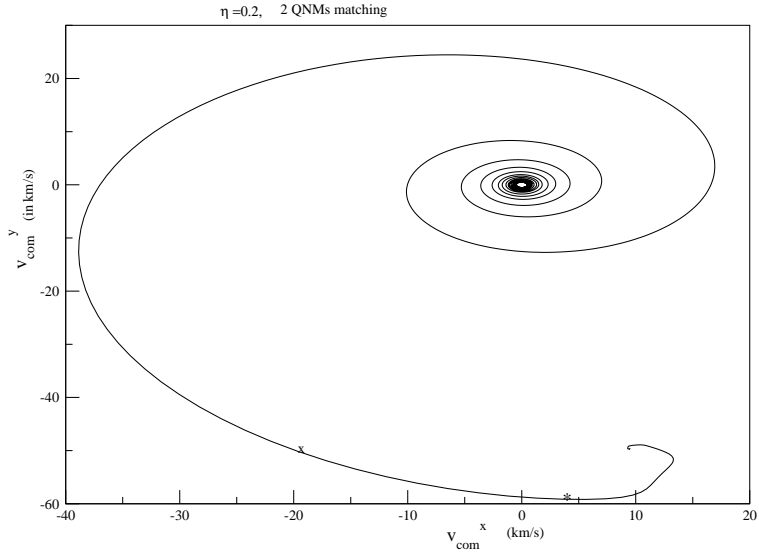}}
\caption{
Parametric plot of $v_{\rm com}^{\rm x}$ versus
$v_{\rm com}^{\rm y}$
during the coalescence of a $\eta =0.2$ binary. We employ the
quasi-Newtonian approximation to the linear momentum flux
followed by a 2 QNMs description of the ring-down phase.
The symbols  $\times $ and $*$ respectively indicate the positions
where the linear momentum flux reached its maximum value and where
matching to the ring-down phase was done.
The initial orbital separation of the binary, when $t=0$, was $r=15$.
\label{Fig_traj}}
\end{center}
\end{figure}

\clearpage

\begin{figure}[c]
\begin{center}
\resizebox{10cm}{!}{\includegraphics{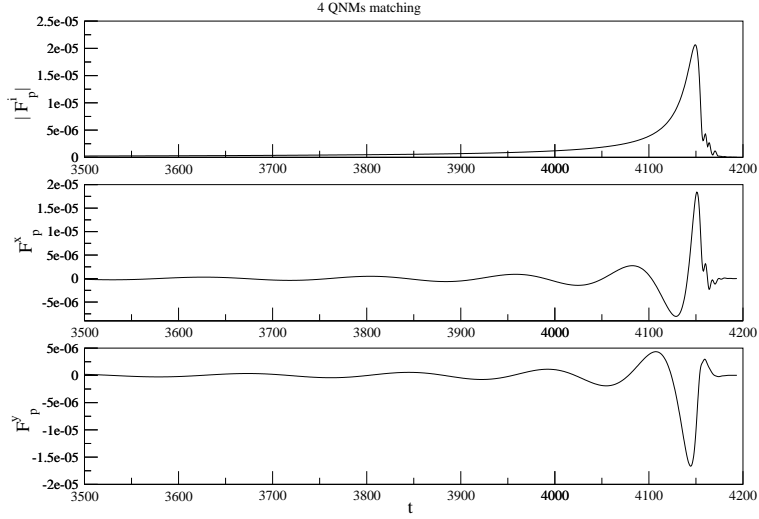}}
\caption{ Plots for the magnitude, $x$ and $y$ components of the
quasi-Newtonian linear momentum flux versus $t$
during the coalescence of a binary with $\eta=0.2$. In this case, the ring-down phase is 
described by 4 QNMs and the initial orbital separation of the binary, when $t=0$,
was $r=15$.
\label{Fig_4_MF_t_B}}
\end{center}
\end{figure}


\begin{figure}[c]
\begin{center}
\resizebox{10cm}{!}{\includegraphics{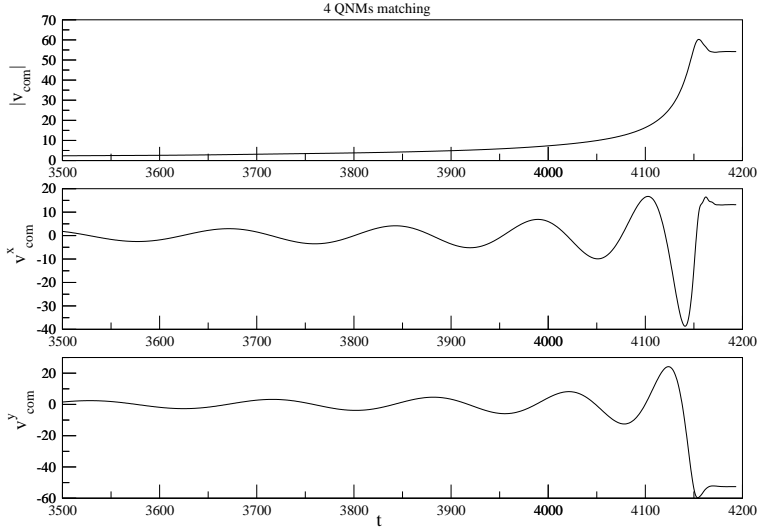}}
\caption{ Temporal plots for the magnitude, $x$ and $y$ components of
recoil velocity for the binary configuration described in 
Fig.~\ref{Fig_4_MF_t_B} ($\eta =0.2$, quasi-Newtonian flux, 4 QNM's).
\label{Fig_5_VC_t_A}}
\end{center}
\end{figure}

\clearpage

\begin{figure}[c]
\begin{center}
\resizebox{8cm}{!}{\includegraphics{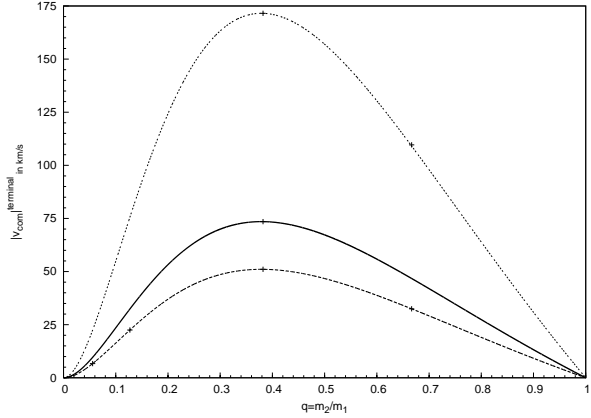}}
\caption{
Our `best-bet' estimate (solid line) for the terminal recoil as a function of the mass 
ratio $q= \frac{m_2}{m_1}$
based on Eq.~(\ref{finalestimate}) with $\tilde F = 1.35$, corresponding
to using the 2PN, Taylor $\tilde F$ in the momentum flux expression
Eq.~(\ref{Eq.19.old}). The ``theoretical uncertainty'' around
this best-bet estimate is illustrated by plotting the results of using:
(i) a quasi-Newtonian flux  (Eq.~(\ref{Eq.19.old}) with
$\tilde F = 1$) [dot-dashed lower curve], or (ii) the Kepler-law-based
2PN flux Eq.~(\ref{BQWflux}) [dotted upper curve].
The estimates taken from Tables~\ref{Tab_1} and \ref{Tab_3} are denoted by $+$ symbols.
\label{Fig_7}}
\end{center}
\end{figure}


\begin{figure}[c]
\begin{center}
\resizebox{8cm}{!}{\includegraphics{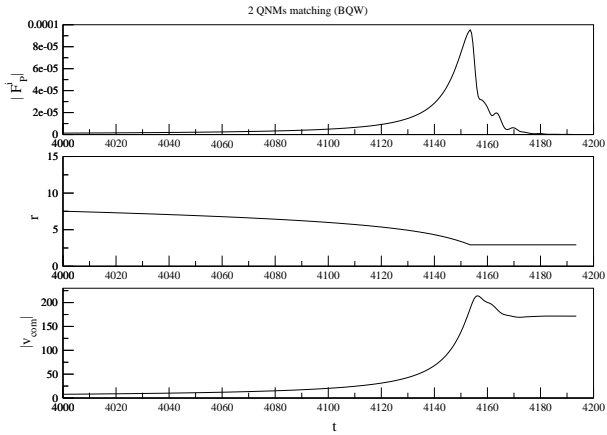}}
\caption{
Temporal plots, along the EOB evolution, for the magnitude of the ``Blanchet-Qusailah-Will''
linear momentum flux, defined by Eq.~(\ref{BQWflux}),
the orbital separation $r$
and the associated recoil $|v_{\rm com}|$ in 
${\rm km/s}$ for $\eta=0.2$ binary.
We terminate the plunge around $r \simeq 3$ and perform two QNMs matching for the ring-down phase. [In Ref.~\cite{BQW05} the evolution was
formally continued down to $r \simeq 2$].
\label{Fig_8}}
\end{center}
\end{figure}

\end{document}